\documentclass[acmsmall]{acmart}

\AtBeginDocument{%
  \providecommand\BibTeX{{%
    \normalfont B\kern-0.5em{\scshape i\kern-0.25em b}\kern-0.8em\TeX}}}

\setcopyright{acmlicensed}
\acmJournal{ACMJCSS}
\acmYear{2023} \acmVolume{1} \acmNumber{1} \acmArticle{1} \acmMonth{1} \acmPrice{15.00}\acmDOI{10.1145/3608115}

\usepackage{amsmath,amsfonts}
\usepackage{algorithmic}
\usepackage{graphicx}
\usepackage{textcomp}
\usepackage{xcolor}
\usepackage{longtable}
\usepackage{appendix}
\usepackage{comment}
\usepackage{booktabs}
\usepackage[para]{threeparttable}
\usepackage{diagbox}
\usepackage{tabularx}
\usepackage{hyperref}

\usepackage{color}
\begin{document}

\title{How Viable are Energy Savings in Smart Homes? A Call to Embrace Rebound Effects in Sustainable HCI}

\author{Christina Bremer}
\email{c.bremer@lancaster.ac.uk}
\orcid{0000-0002-8222-9786}
\affiliation{
  \institution{Lancaster University}
  \city{Lancaster}
  \country{United Kingdom}
}

\author{Harshit Gujral}
\email{harshit@cs.toronto.edu}
\orcid{0000-0001-5152-1971}
\affiliation{
  \institution{University of Toronto}
  \city{Toronto}
  \country{Canada}
}

\author{Michelle Lin}
\email{michelle.lin2@mail.mcgill.ca}
\orcid{0009-0003-4476-5558}
\affiliation{
  \institution{McGill University}
  \city{Montreal}
  \country{Canada}
}

\author{Lily Hinkers}
\email{lily.hinkers@campus.tu-berlin.de}
\orcid{0009-0007-9120-9799}
\affiliation{
  \institution{Technische Universität Berlin}
  \city{Berlin}
  \country{Germany}
}

\author{Christoph Becker}
\email{christoph.becker@utoronto.ca}
\orcid{0000-0002-8364-0593}
\affiliation{
  \institution{University of Toronto}
  \city{Toronto}
  \country{Canada}
}

\author{Vlad C. Coroam\u{a}}
\email{coroama@tu-berlin.de}
\orcid{0000-0002-9292-3886}
\affiliation{
  \institution{Technische Universität Berlin}
  \city{Berlin}
  \country{Germany}
}

%% By default, the full list of authors will be used in the page headers. Often, this list is too long, and will overlap other information printed in the page headers. This command allows the author to define a more concise list of authors' names for this purpose.

%% The abstract is a short summary of the work to be presented in the article.
\begin{abstract}
As part of global climate action, digital technologies are seen as a key enabler of energy efficiency savings. A popular application domain for this work is smart homes. There is a risk, however, that these efficiency gains result in \textit{rebound effects}, which reduce or even overcompensate the savings. Rebound effects are well-established in economics, but it is less clear whether they also inform smart energy research in other disciplines. In this paper, we ask: to what extent have rebound effects and their underlying mechanisms been considered in computing, HCI and smart home research? To answer this, we conducted a literature mapping drawing on four scientific databases and a SIGCHI corpus. Our results reveal limited consideration of rebound effects and significant opportunities for HCI to advance this topic. We conclude with a taxonomy of actions for HCI to address rebound effects and help determine the viability of energy efficiency projects.
\end{abstract}

%% The code below is generated by the tool at http://dl.acm.org/ccs.cfm. Please copy and paste the code instead of the example below.



%% Keywords. The author(s) should pick words that accurately describe the work being presented. Separate the keywords with commas.

\received{16 February 2023}
\received[revised]{1 June 2023}
\received[accepted]{8 June 2023}

%% This command processes the author and affiliation and title information and builds the first part of the formatted document.
\maketitle

\section{Introduction}

One of the United Nations Sustainable Development Goals is \emph{climate action}. As energy consumption is tightly linked to the emission of greenhouse gases---which effect climate change---energy reductions are considered an important lever in our mitigation efforts. And they are urgently needed: the Sustainable Development Goals Report 2022 specifies that energy-related CO\textsubscript{2} emissions increased by 6\% in 2021 to the highest ever level~\cite{UNGoalsreport}. 

To drive climate action, digital technologies have frequently been employed for energy efficiency purposes. Within HCI and the computing community more broadly, a popular application domain for this work is smart homes. Defined as interconnected and Internet-enabled residential buildings that are equipped with automation technology, smart homes have sparked increasing research interest. From a sustainability perspective, their automation technology ``promises considerable savings of energy, therefore, simultaneously reducing the operational costs of the building over its whole lifecycle''~\cite{reinisch2011thinkhome}. Considering the carbon impact of the built environment, these savings could be vital in the context of climate change mitigation: according to the United Nations Environment Programme, the buildings and construction sectors are responsible for 36\% of the global final energy consumption and 37\% of the global energy-related CO\textsubscript{2} emissions~\cite[Fig. 2]{UNreport}. In turn, a significant proportion of these estimates (61\% and 49\%, respectively, amounting to 22\% and 17\% of the total energy and energy-related emissions) can be attributed to residential energy consumption~\cite[Fig. 2]{UNreport}, and particularly to heating, ventilation and air conditioning~\cite{perez-lombard_review_2008}. 

While, in principle, energy efficiency promises significant savings, it also comes with an energy footprint of its own (from e.g., hardware infrastructure and upgrading of devices due to software updates \cite{strengers2016hidden}) and regularly causes~\emph{rebound effects}~\cite{a_greening_energy_2000}. The term was originally introduced to describe a phenomenon in the energy market, i.e., that more efficient steam engines did not lead to less coal consumption, but---as their application domains were expanding together with their  increasing efficiency---to ever \textit{more} consumption of coal. Today, the term `rebound effect’ is applied more widely. In the original energy context, rebound effects refer to an array of mechanisms that counteract the energy-saving potential of energy efficiency measures~\cite{sorrell_jevons_2009}; the definition, however, can be extended to cover additional forms of resources. For example, the idea of expanding highways to cope with traffic yields a short-term relief but very often results in \textit{more} traffic in the long run---a phenomenon also known as \textit{generated traffic}~\cite{duranton2009fundamental}. Similarly, when a technology induces time savings, time saved is then used to perform energy-intensive activities~\cite{binswanger_technological_2001}. When the rebound effect is moderate, the overall balance might still be beneficial, leading to an overall energy or resource decrease. The specific case in which the rebound effect is larger than the initial gains, leading to an overall energy consumption~\emph{increase}, is often referred to as `Jevons' Paradox’ or `Khazzoom-Brookes postulate’~\cite{sorrell_jevons_2009}. 

For (smart home) energy efficiency, rebound effects mean that when energy efficiency improves, energy demand may decrease less than anticipated or even increase. Importantly, this phenomenon stems from the interplay of technical characteristics with social, psychological, economic and cultural factors~\cite{Strengers2011, Schiano2010, tirado_herrero_smart_2018}; this makes modelling approaches challenging and calls for situated research. At the same time, rebound effects raise questions about the large set of persuasive, awareness-enhancing, or efficiency-focused research in the field \cite{disalvo2010mapping}. While the HCI community has a history of deep and critical engagement with smart energy and demand reduction, especially the expectations, experiences and attitudes of energy consumers \cite{strengers2014smart,jensen2018designing,hargreaves2010making,strengers2019protection}, it is unclear to what extent they have engaged with the concept of rebound effects and how can they effectively apply their skills to deepen our understanding of it. At CHI 2011, Kaufman and Silberman called the HCI community to broaden their evaluations beyond simple proxies and consider rebound effects~\cite{kaufman2011rebound}. Has this call been heard?

Rebound effects are a tangible, well-established concept in the energy literature~\cite{hens2010energy,sorrell_jevons_2009}. While the concept is not native to computing and HCI research, it affects the conclusions we can draw from their studies and, thus, the design of technology in the energy space. So if we do not consider these effects, we risk the research becoming misdirected, and its efficiency savings might backfire. If energy research informs decisions on a policy level, we also risk that policy makers direct their funding and reforms towards efficiency measures which ultimately may not bring the anticipated results. In this context, we believe that the HCI community could take on an important, pioneering role due to its research focus on smart homes and expertise on how to holistically study socio-technical phenomena, such as rebound effects.

To explore the attention paid to rebound effects within the computing community, and HCI community more specifically, this study investigates the prevalence of rebound effect considerations in energy efficiency research. We focus on \textit{smart home} research as a representative domain in which energy efficiency research is prominently placed. To achieve this, we conduct a systematic literature mapping in four scientific databases for possible combinations of the keyword clusters `smart home’, `energy efficiency’ and `rebound effect’. Through this mapping, we first aim to establish to what extent smart home energy efficiency research---within and beyond HCI---has adopted the concept of rebound effects and how this compares to rebound effect considerations in energy efficiency research more generally. In a second step, we outline the benefits of embracing rebound effects as a concept in HCI energy research and show how the community can effectively apply their skills and expertise to advance energy efficiency efforts. 

\section{Background} \label{sec:background}

Defined as residential buildings that are equipped with systems and devices that are Internet-enabled and interconnected and that can be automatically and remotely controlled~\cite{li_jiang_smart_2004}, smart homes have been an active area of energy research for over three decades \cite{lutolf1992smart}. Due to their energy-saving potential, smart homes are one of the European Union's ten priority action areas in its Strategic Energy Technology Plan~\cite{ECreport}. In line with this prioritisation, numerous sources identify quite substantial energy-saving potential for smart homes: deploying in-home displays (IHDs) to continuously reveal the overall electricity consumption of a smart home to its inhabitants has been the focus of various studies, and was found to yield electricity savings of between 0\% and 18\%~\cite{ford_categories_2017}. Looking beyond electricity and considering the more substantial energy consumption for space heating, Becker et al.~\cite{becker_estimating_2018} find an average energy-saving potential of over 9\% in a smart heating simulation based on occupancy detection and prediction from real-world data of more than 3000 Irish households. Considering all sources of energy consumption in a household, Williams and Matthews~\cite{williams_scoping_2007} suggest that monitoring and control systems can help occupants lower their energy consumption by 3---26\%. While the lower end of this estimate applies to the use of programmable thermostats, the upper end refers to ``an integrated system including monitoring and control of appliances, plus zone heating/cooling''. This is consistent with the claim that the energy-saving potential of smart energy systems tends to be positively linked to the overall level of complexity, with the cost of greater computational requirements~\cite{naylor_review_2018}. However, as neither rebounds nor the energy footprint of these complex monitoring and sensing devices are consistently accounted for, their impact on the estimated saving potential remains to be determined.

\subsection{Challenging Efficiency Visions for Smart Homes}

Realising the energy-saving potential of smart homes can be challenging: Darby~\cite{darby_smart_2018} points out the scarce number of evaluations under real-world settings in favour of simulations and evaluations under controlled laboratory conditions. This imbalance can be problematic and lead to an instantiation of the well-documented energy performance gap between ex-ante energy saving predictions and ex-post assessments~\cite{gram-hanssen_energy_2018}, due to e.g., contextual factors, building and installation design parameters, and occupant behaviour not being taken into account~\cite{van_thillo_potential_2022}. In 2018, Hargreaves and Wilson~\cite{hargreaves_learning_2018} reported the results from a study of smart home technology in the real world: based on a 9-month field project with ten UK households, they found that the technology can even lead to overall \emph{energy intensification} through, e.g., increased comfort expectations and pre-warming of domestic spaces. Additional challenges that may impact the energy-saving potential of smart homes include their limited uptake (due to unawareness or inaccessibility~\cite{williams_scoping_2007,tirado_herrero_smart_2018} and rebound effects. Tirado Herrero et al. \cite{tirado_herrero_smart_2018} conclude that ``in contrast to aspirational claims for a `smart utopia' of greener, less energy-intensive, and more comfortable homes [...] SHTs [smart home technologies] may reinforce unsustainable energy consumption patterns in the residential sector''.

The HCI community specifically has a longstanding record of work that talks in nuanced ways about how smart homes can undermine energy savings by introducing new lifestyle expectations and changing behaviours. This includes the identification of socio-technical challenges around integrating new technologies into everyday routines and practices~\cite{costanza2014doing,jakobi2017catch} and the often energy-intensive visions for smart homes. Commonly associated with concepts such as desirability, convenience (including effortless energy savings) and control, these concepts can undermine the sustainability promises of smart homes~\cite{jensen2018exploring,strengers2020pursuing}. Gendered and cultural expectations~\cite{strengers2019protection,dankwa2020investigating} and individual preferences~\cite{jensen2018designing} add another layer of complexity, but are not consistently acknowledged. As Strengers~\cite{strengers2014smart} points out, many studies have focused on ``resource man'', a default male technology-savvy energy consumer who makes rational decisions about energy use. 

From an energy justice perspective, smart energy technology can also be problematic. Research finds that, despite their desire to monitor and regulate their energy usage, many low-income households do not receive feedback on their consumption~\cite{dillahunt2009s}; a problem that is reinforced through e.g., the high cost of sensing and feedback technologies that can support energy savings~\cite{kallman2019power,milchram2018energy}. Calls to provide easier and subsidised access to these technologies~\cite{podgornik2013impact} are thereby potentially in tension with concerns about data privacy and security that relate to their widespread use~\cite{kallman2019power,milchram2018energy}. Working class people might also end up paying more for their energy \textit{despite} having e.g., smart meters installed as they cannot adjust their schedule to use energy when it is cheaper~\cite{kallman2019power}. In this context, it will be critical to make energy feedback and energy-saving support accessible and affordable across society. At the same time, care must be taken not to rely on smart technologies and smart homes as the solution. Our work specifically focuses on the limitation of rebound effects (as one of the reasons to challenge this reliance) and calls for a more socio-technical understanding of these effects. This understanding can help pave the way for policies that reduce the ineffective application of efficiency technologies while also reducing inequalities.

\subsection{Rebound Effects} \label{sec:retypes}

The umbrella term `rebound effects' describes a rich class of mechanisms occurring when goods or services become easier accessible or more attractive, leading to their increased demand, therefore their increased consumption and production, and thus ultimately increased energy consumption, material consumption, and greenhouse gas emissions. The first such mechanism yielding rebound effects was described as early as 1865 by British economist William S. Jevons. In his book ``The Coal Question'', Jevons notices that although coal extraction was becoming ever more energy efficient at the time, overall not less but more energy was used in coal extraction~\cite{jevons1865coal}. The efficiency gains of coal extraction were thus a victim of their own success: by making coal more affordable, they were making it attractive for an increasing number of applications and customers, while at the same time making the process of coal extraction more scalable. 

\subsubsection{Direct Rebound and Backfire} \label{sec:rebound-direct}

The rebound effect that increases the attractiveness of the very good that has been more efficiently produced is today often addressed as the `direct rebound effect'~\cite{coroama_digital_2019}. After being largely ignored for more than a century, it was brought back to scientific scrutiny by Khazzoom in 1980~\cite{khazzoom1980:rebound}. Neither Jevons nor Khazzoom 100 years later used the term `rebound effect' (although several sources wrongly claim the latter coined the term), which started to be widely used around the year 2000, e.g.,~\cite{a_greening_energy_2000, saunders2000:rebound, berkhout_defining_2000}. Rebound effects are generally expressed in relation to the direct savings that triggered them in the first place: when smaller than 100\%, rebound effects reduce these savings but there is still an overall resource reduction, when they reach 100\%, they fully compensate the initial savings, and when larger than 100\%, they outweigh them. In many instances, rebound effects do indeed reach 100\% or more. For electric lighting, for example, a large-scale study came to the conclusion that ``global energy use for lighting has experienced 100\% rebound over 300 years, six continents, and five technologies''~\cite{tsao_solid-state_2010,saunders_rebound_2012}.

According to their empirical observations, both Jevons~\cite{jevons1865coal} and Khazzoom~\cite{khazzoom1980:rebound} postulated the size of the rebound effect to always be larger than that of the original savings, thus not only reducing the original savings but outweighing them. As noted by both Alcott~\cite{alcott2005:jevons} and Sorrell~\cite{sorrell_jevons_2009}, `postulate' is the correct wording in this context, as there is not enough evidence to support the claim that rebound always exceeds 100\%, and increasing evidence points to examples where it does not~\cite{gillingham2013:rebound-overplayed}. The particular case, in which efficiency gains paradoxically led to more overall consumption, is thus known in the literature as `Jevons' paradox'~\cite{alcott2005:jevons} or `backfire' (unlike `rebound', the term `backfire' was indeed coined by Khazzoom~\cite{khazzoom1980:rebound}). To reflect another economist's (Len Brookes) groundbreaking work in the field (e.g., \cite{brookes1990:rebound-ghg}), backfire has also been called `Khazzoom-Brookes postulate'~\cite{saunders1992:khazzoom-brookes}.

\subsubsection{Indirect and Structural Rebound, and their Many Flavours}
\label{sec:rebound-indirect}

Efficiency gains and associated price reductions can have various consequences. They do not necessarily (or at least not exclusively) lead to more demand for the very good that is being more efficiently produced. Monetary savings from the more efficient production of a good or service can be spent on different energy- and resource-consuming activities, for example---this is known as `income effect'~\cite{berkhout_defining_2000}. Additionally, a lower price makes a product relatively more affordable than other, similar products, which it may subsequently partly substitutes---a phenomenon known as `substitution effect'. The increased consumption can thus propagate to other goods and sectors~\cite{berkhout_defining_2000}, making income and substitution effects prime examples of what is often called `indirect rebound effects'~\cite{coroama_digital_2019}. 

Energy efficiency, in particular, can have a catalyst effect on productivity, boosting total factor productivity and leading to cascading rebound effects throughout the economy. Higher productivity can thus lead to more overall economic output, which in turn requires more (energy) inputs~\cite{sorrell_jevons_2009}, and various readjustments along the entire economy. These macroeconomic effects, which are conceptually hard to grasp and almost impossible to quantify, can be referred to as `structural rebound', `macroeconomic rebound'~\cite{barker2009:macroeconomic-rebound} or `transformational' rebound~\cite{pohl2019:indirect-lca,borjesson_rivera2014}; they are often included in a larger category of `systemic transformations' brought about by digitalisation~\cite{williams2011:indirect-effects}. 

Beyond efficiency gains (and thus energy or material savings), several other types of savings can represent the trigger for rebounds. In particular, rebound effects can emerge as a consequence of decreased transaction costs (i.e., monetary or non-monetary costs to access a good, evaluate it, and general market information), due to time savings~\cite{binswanger_technological_2001} (leading to `time rebound'), or even due to the decreased skills required to perform an action such as driving a car~\cite{coroama_skill_2020} (yielding `skill rebound'). Given this variety of subtle effects and influence mechanisms, there is no generally agreed-upon taxonomy for indirect rebound effects. What distinguishes them from direct rebound effects, however, is that they either lead to an increase of consumption, but for different products (and not the one being initially more efficiently produced), or through other means and not as a consequence of a price reduction (but for example through time or transaction costs savings)~\cite{coroama_skill_2020}.

\subsection{Accounting for Rebound Effects in Energy Research}

In research fields such as energy policy, rebound effects are a firmly established concern~\cite{safarzadeh2020review,wang2018evaluation}. As the strength of the rebound effect varies significantly across contexts, a key research goal in these fields is to identify causal factors that allow a concrete, context-specific estimation of rebound effects, to then direct and focus possible intervention points~\cite{a_greening_energy_2000}. By understanding causal relationships and context-specific factors, it is argued that energy-focused research can be targeted and effective. The methods used to acquire this understanding include modelling and simulation~\cite{wang_measurement_2016,nassen_quantifying_2009} as well as empirical studies~\cite{rau_influence_2018,haas_rebound_2000} that take both macro-level policy perspectives and micro-level household perspectives. Some of these studies specifically highlight the need to consider rebound effects in smart homes. For example, one study of the rebound effect in smart homes suggests that ``the rebound effect on energy use should not be underestimated''~\cite{rau_influence_2018}. It is important to understand what causes rebound effects, and how large they are, to direct research priorities~\cite{wang_measurement_2016}. For all these reasons, researchers should give ``due consideration to the full range of behavioural responses to technical efficiency''~\cite{a_greening_energy_2000} when conducting energy-focused research on smart homes. After all, the actual energy consumption of a smart home is not determined by its technical components alone but must be understood as a socio-technical system.

A substantial part of the literature on smart homes, however, originates in engineering sciences. And although it has been shown that digital technologies are particularly prone to rebound effects---leading even to the emergence of the new term `digital rebound'~\cite{coroama_digital_2019}---it has not yet been thoroughly analysed to which extent the computing communities at large, and that on smart homes, in particular, have engaged with the concept of rebound effects.

In 2011, Kaufman and Silberman~\cite{kaufman2011rebound} encouraged the sustainable HCI community to consider rebound effects to mitigate the problem that ``proxies generally used in evaluation (e.g. less energy or water consumption) may be poor indicators of such systems’ effect on emissions because of effects outside the scope of analysis''. At this point, the authors were ``not aware of studies examining the degree to which the effects of sustainable HCI technologies rebound''. Assessing whether this has changed over the past decade, and how the level of perception within HCI compares to the one rebound effects receive in computing generally and, even broader, in all energy efficiency-related fields, has been a key motivation behind this work.

\section{Methodology}

\begin{figure}
\centerline{\includegraphics[width=6cm]{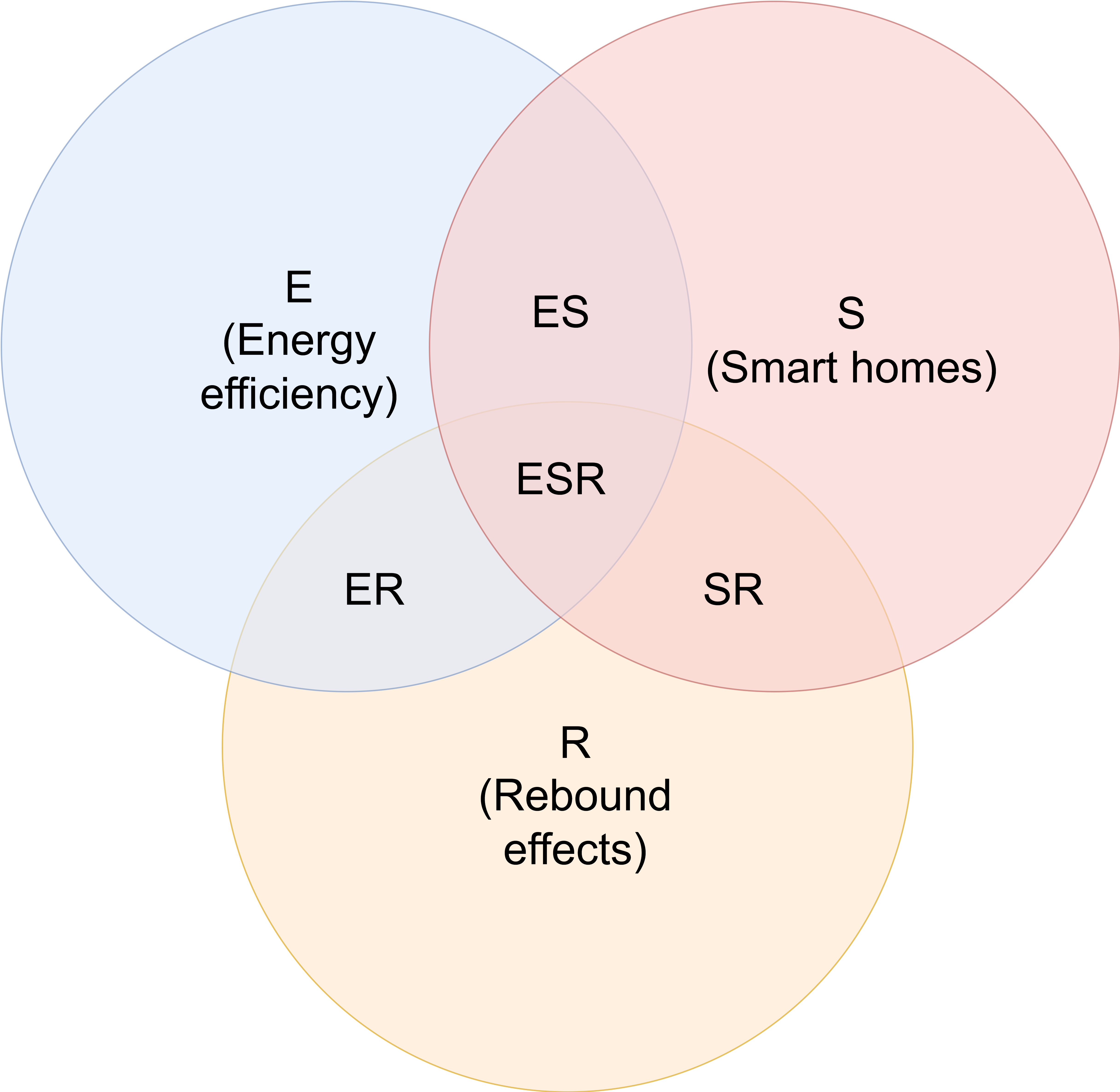}}
\caption{Intersections of energy efficiency, smart homes, and rebound effects.}
\label{fig:venn}
\Description{A Venn diagram depicting sets for ‘Energy efficiency’ (denoted by a capital E), ‘Smart homes’ (denoted by a capital S), and ‘Rebound effects’ (denoted by a capital R). The 3 pair-wise intersections are correspondingly noted ES, ER, and SR, respectively, while the intersection of all three sets is denoted ESR.}
\end{figure}

We evaluate to what degree the work in a range of fields considers rebound effects, in order to evaluate what energy efficiency research in HCI could potentially learn from and contribute to other fields. To determine the coverage of rebound effects in smart home energy efficiency research, we identified the publications in the following three research domains and in their intersections: smart homes, energy efficiency, and rebound effects. The intersections are visualised in the Venn diagram in Fig.~\ref{fig:venn}. This mapping of sets allows us to identify which publications attend to rebound effects in smart home research, evaluate the relative frequency of attention in contrast to their larger research areas, and compare the attention to rebound effects in smart home research with more established research fields examining energy efficiency and policy. We conducted searches for literature in each of these areas in multiple databases: general scientific databases (i.e., Scopus and ScienceDirect), general computing ones (i.e., IEEE and ACM), and HCI-specific venues (i.e., SIGCHI-sponsored conferences as well as the PACMHCI journal, as described below).

To identify the HCI-specific venues, we proceeded as follows: from SIGCHI's ``upcoming conferences'' website\footnote{https://sigchi.org/conferences/upcoming-conferences/}, we chose all 2022 conferences, which are 100\% sponsored by SIGCHI: CHI itself, 11 specialised conferences, and 14 in-cooperation conferences (26 in total). From these, we then excluded 6, because we could not identify their collection of proceedings (CI, HUCAPP, MuC), their respective proceedings' website were not standalone searchable, but a search would yield results from the entire ACM full paper collection (LAK, AM), or were not English-speaking (IHM). Our corpus, which we refer to as the `SIGCHI library', consisted of the all-time proceedings for the remaining 20 conferences (CHI, TEI, C\&C, DIS, IMX, EICS, AutomotiveUI, RecSys, MobileHCI, ICMI, CSCW, AHs, CHIIR, W4A, GI, AVI, CUI, iWOAR, NordiCHI, and IoT) plus all volumes of the PACMHCI journal.

Systematic mapping studies are a form of literature review that focuses on structuring a research area rather than gathering and synthesizing all evidence~\cite{petersen2008systematic,petersen_guidelines_2015}. While the synthesis focus of targeted reviews is more appropriate for questions that allow the aggregation and comparison of findings across different studies~\cite{kitchenham_systematic_2009,azevedo2014consumer}, mapping studies can be very useful to identify trends and opportunities for research spanning multiple fields. In our case, we are especially interested in structuring research on our subject by identifying what is published at the outlined intersections. We will then review the evidence in the most relevant intersections in more depth.

\begin{figure*}
\begin{centering}
\centerline{\includegraphics[width=\linewidth]{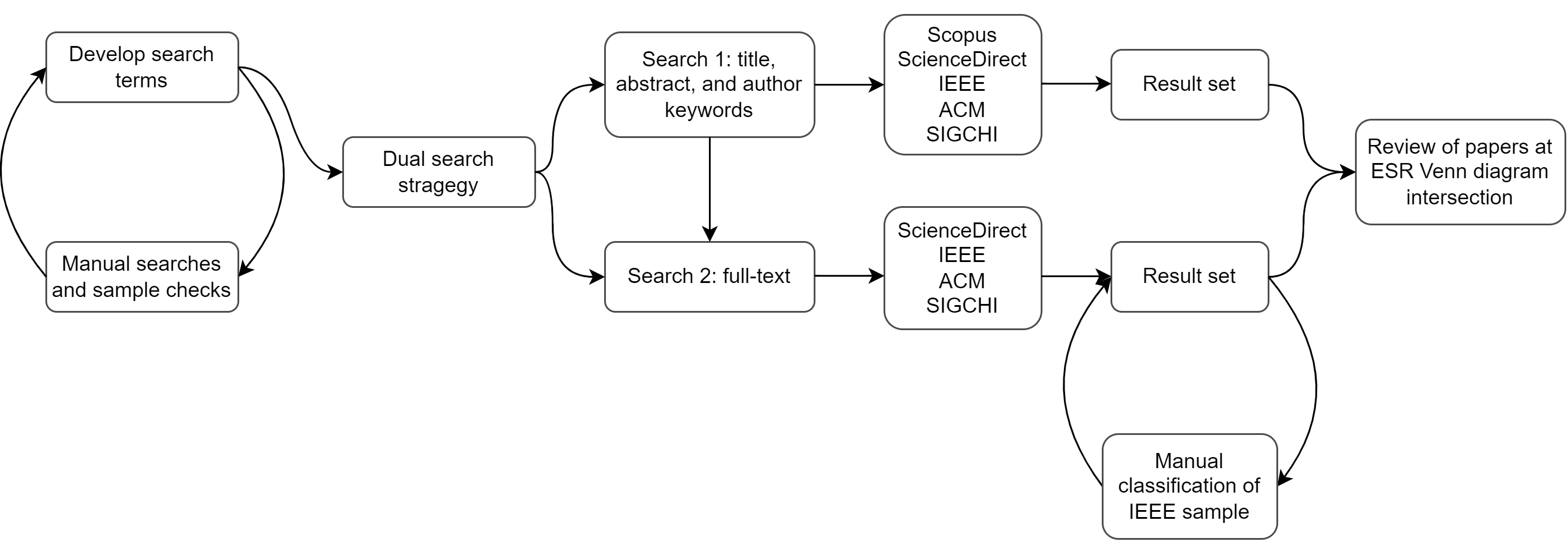}}
\caption{A flowchart showing each step of the literature mapping process.}
\label{fig:method}
\Description{A flowchart showing the individual steps of the literature mapping process. On the very left, there is a cycle between ‘develop search items’ and ‘manual searches and sample checks’. After this fine-tuning was ready, an arrow exits the cycle leading to ‘dual search strategy’. The two strategies presented on two parallel tracks: the upper track starts with ‘Search 1: title, abstract, and author keywords’ in the five databases Scopus, ScienceDirect, IEEE, ACM as well as the SIGCHI subset. The lower track starts with ‘Search 2: full text’ in the four databases where it was possible: ScienceDirect, IEEE, ACM, and the SIGCHI subset. Each of the tracks results in a ‘Results set’ final step, while the result box of the lower set has a cycle with ‘manual classification of IEEE sample’. Finally, from each result set, the two tracks are brought back together into ‘review of papers at ESR Venn diagram intersection’.}
\end{centering}
\end{figure*}

The literature search and mapping workflow is described in Fig. \ref{fig:method}. The search terms were evaluated and tuned by conducting manual searches and sample checks to ensure relevance. Our final search terms used the following keywords and combinations for the three areas we cover (the discussion in Section~\ref{sec:rebound-direct} motivates the choice of search terms for rebound effects):

\begin{enumerate}
    \item ``smart home*''
    \item ``energy efficien*''
    \item ``rebound effect*'' OR ``Jevons paradox'' OR ``Khazzoom-Brookes postulate''
\end{enumerate}

Because the ACM Guide to Computing Literature and ScienceDirect do not allow the use of wildcards as placeholders within quotation marks, the keywords/keyword combinations used there were as follows: 
\begin{enumerate}
    \item ``smart home'' OR ``smart homes''
    \item ``energy efficient'' OR ``energy efficiency''
    \item ``rebound effect'' OR ``rebound effects'' OR ``Jevons paradox'' OR ``Khazzoom-Brookes postulate''
\end{enumerate}

In a second step, we used the search terms to conduct searches in four scientific databases, namely Scopus, ScienceDirect, the ACM Full-Text Collection and IEEE Xplore. These databases were chosen to cover a representative proportion of archived literature within and beyond traditional computer science disciplines. Our searches were carried out between December 2021 and April 2022. Notably, there is no start date for our searches, implying they include all the articles that were indexed in the database. For instance, searching ``energy efficien*'' on IEEE Xplore yields articles from as early as 1889. We used a dual search strategy, implying that we conducted our searches twice: (1) limited to the publications’ titles, abstracts and keywords, referred to as \textit{Search 1}, and (2) full-text searches (i.e., searching for the keywords anywhere in the publications), referred to as \textit{Search 2}. The intention for \textit{Search 1} was to maximise the likelihood that the returned publications exhibit a focus on smart homes, energy efficiency and/or rebound effects, rather than to merely mention these terms in passing within a different context. In IEEE, titles, abstracts and keywords could not be selected as a set of fields due to a technical constraint, so the search covers the closest equivalent: ``all metadata''. The detailed list of search terms can be found in the supplementary material; the scripts used to confirm the findings for Scopus and ScienceDirect from manual searches and to establish the recent trends described below can be found here: \url{https://doi.org/10.5281/zenodo.8062778}. 

The metadata-focused \textit{Search 1} may in principle omit papers that cover rebound effects but do not mention them in their title, abstract or keywords. This potential gap is addressed by combining the first search strategy with the second---which also enabled us to compare the number of search results for papers that focus on rebound effects (\textit{Search 1}) and those that mention them but might not prioritise them (\textit{Search 2}). Searching only in titles, abstracts and keywords \textit{(Search 1)} might exclude publications that would thematically fit our search, so we intended the full-text searches \textit{(Search 2)} as a strategy to rule out that the exclusion would skew the results. As Scopus does not store full texts, this was only possible for ScienceDirect, IEEE, ACM and SIGCHI.

Since our selected databases represent overlapping but distinct academic disciplines, we computed percentages to compare the considerations of rebound effects among these databases. In particular, we calculated the following two ratios for \textit{Search 1} and \textit{Search 2}: (1) rebound effect considerations in the energy efficiency domain in general, and (2) rebound effect considerations in the energy efficiency domain for smart homes. The rebound effect considerations in the energy efficiency domain were calculated by, 

\begin{equation}
    {R \: in \: E} \: (\%) = (E  \cap R) * 100 / E
\end{equation}

where $E$ represents the number of papers in the energy efficiency domain, and $E \cap R$ denotes the number of papers in the energy efficiency domain that consider rebound effects. Similarly, rebound effect considerations in the energy efficiency domain for smart homes were expressed as

\begin{equation}
    {R \: in \: E \: for \: S} \: (\%) = (E  \cap R \cap S) * 100 / (E \cap S)
\end{equation}

where $E \cap S$ represents the number of papers in the energy efficiency and smart homes domain, and $E \cap S \cap R$ denotes the number of papers at the intersection of energy efficiency and smart homes that pay attention to rebound effects.  
To refine and evaluate the search strategy, including the search terms, we conducted manual checks on the IEEE result sets, with emphasis on the number of ‘false positives’ returned in our full-text searches, i.e., the number of returned papers that do not focus on smart homes, energy efficiency and/or rebound effects. As a database focused on engineering and computing, IEEE is highly relevant for smart home research and its result set was considered a representative sample for the research goals. To understand the development in engagement with rebound effects over time, we conducted a temporal analysis of the smart homes energy efficiency research for \textit{Search 1} and \textit{Search 2}. We conducted this analysis for a ten-year period from 2011 to 2020 across all selected databases. It should be noted that we excluded the year 2021 because these databases often report a lag in indexing recent literature.

Finally, to understand and learn from the existing literature on rebound effects in smart home energy efficiency, we conducted a detailed review of the papers that contain all keywords (area ESR in the Venn diagram in Fig. \ref{fig:venn}) in their titles, abstracts or keywords. For computing specifically (i.e., IEEE and ACM), there were no papers comprising all keywords in \textit{Search 1} (and thus for ACM's SIGHCI subset a fortiori not), we performed the same detailed review on the SIGCHI papers containing all three keywords anywhere in the full text (i.e., for \textit{Search 2}). 

\section{Results}

Here we first present the quantitative results of all searches and their temporal evolution, and then discuss and compare the attention to rebound effects within computing and in the broader scientific literature.

\subsection{Overview}

Energy efficiency is a vast area of research across all fields, with Scopus returning almost a quarter-million papers in \textit{Search 1} alone. In ScienceDirect and IEEE, the numbers are slightly lower but still substantial with around 40,000 and 66,000 search results for `energy efficiency’ respectively. In ACM's full text database, there are significantly fewer results (almost 7,000), while for the SIGCHI library, energy efficiency is clearly not a crucial aspect, with only 33 results in title, abstract, or keywords. Smart home research is more evenly distributed, with approximately 15,000, 1,000, 7,000, and 3,000 papers appearing in \textit{Search 1} for the four large databases, and 200 even in ACM's SIGCHI subset. By contrast, the number of papers mentioning `rebound effects’ is limited to a couple thousand at most, almost all of which appear outside of the computing databases: Scopus, ScienceDirect, IEEE and ACM contain 2,714, 1,114, 79, and 7 papers, respectively, and SIGCHI none whatsoever (see Table \ref{tab:search1}, row 3). 

\begin{table*}
\centering
\begin{threeparttable}[b]
\small
\renewcommand{\arraystretch}{1.5}
\caption{Results from Search 1: titles, abstracts or author keywords}
\centering
\begin{tabular}{cp{0.5cm}p{1.2cm}p{0.7cm}p{1.1cm}p{1.5cm}p{1.7cm}p{1.0cm}p{1.0cm}p{1.0cm}}

\toprule
\textbf{Row} & \textbf{Set} & \textbf{Energy \newline efficiency} & \textbf{Smart home} & \textbf{Rebound effects} & \textbf{Scopus} & \textbf{ScienceDirect} & \textbf{IEEE} & \textbf{ACM} & \textbf{SIGCHI} \\ \midrule

1 & E & X & && 249,167 & 39,663 & 66,196 & 6,842 & 33 \\\hline
2 & S & & X && 15,502 & 1,113 & 7,070 & 1,149 & 200 \\\hline
3 & R & && X & 2,714 & 1,114 & 79 & 7 & 0 \\\hline
4 & ES & X & X& & 895 & 93 & 367 & 47 & 3 \\\hline
5 & ER &  X && X & \textbf{607 (0.24\%)\tnote{*}} & \textbf{337
(0.84\%)\tnote{*}} & \textbf{17 (0.02\%)\tnote{*}} &\textbf{1 (0.01\%)\tnote{*}} & \textbf{0} \\\hline
6 & SR & & X & X & 5 & 1 & 1 & 0 & 0 \\\hline
7 & ESR & X & X & X & \textbf{4 (0.45\%)\tnote{**}} & \textbf{1 (1.07\%)\tnote{**}} & \textbf{0} & \textbf{0} &  \textbf{0} \\\bottomrule

\end{tabular}
\begin{tablenotes}
       \item [*] as per Eq. 1
       \item [**] as per Eq. 2
     \end{tablenotes}
\label{tab:search1}
\end{threeparttable}
\end{table*}

Strikingly, not a single paper in the field of computing (neither IEEE nor ACM) simultaneously encompasses all three topics based on \textit{Search 1}. Scopus and ScienceDirect yield the small but non-trivial numbers of 4 papers and 1 paper, respectively (see Table \ref{tab:search1}, row 7). Proportionally, the highest percentage of papers on `energy efficiency' that mention `rebound effects' can be found in Scopus (0.24\%) and ScienceDirect (0.84\%), covering several hundred publications; with less than 1\%, the percentages are still quite low. Not remotely as low, however, as for the computing literature. As Table~\ref{tab:search1} shows, for general energy efficiency, the ratio of papers focusing on rebound effects is only 0.02\% and 0.01\% for IEEE and ACM, respectively---one to two orders of magnitude lower than in the two general literature databases.

In principle, this could be an artefact of emphasis: perhaps computing papers engage with rebound effects but consider them important only within the context of discussing work in the body of text? But the results from \textit{Search 2} show that this is unlikely. As Table \ref{tab:search2} documents, they are broadly consistent with the findings above. For instance, in the thousands of ScienceDirect papers on `energy efficiency', 1.7\% mention `rebound effect'. In contrast, IEEE and ACM yield fewer papers---251 (0.14\%) and 43 (0.23\%), respectively (see Table \ref{tab:search2} row 5). 

\begin{table*}
\centering
\begin{threeparttable}[b]
\small
\renewcommand{\arraystretch}{1.5}
\caption{Results from Search 2: full-text}
\begin{tabular}{cp{0.5cm}p{1.4cm}p{0.9cm}p{1.2cm}p{1.8cm}p{1.7cm}p{1.2cm}p{1.3cm}}

\toprule

\textbf{Row} & \textbf{Set} & \textbf{Energy \newline efficiency} & \textbf{Smart home} & \textbf{Rebound effects} & \textbf{ScienceDirect} & \textbf{IEEE} & \textbf{ACM} & \textbf{SIGCHI} \\\midrule
1 & E & X & && 212,029 & 170,243 & 18,928 & 278 \\\hline
2 & S & & X && 7,641 & 25,977 & 5,078 & 777 \\\hline
3 & R &&& X & 11,779 & 797 & 98 & 23 \\\hline
4 & ES & X & X && 2,540 & 5,544 & 694 & 41\\\hline
5 & ER & X && X & \textbf{3,612 (1.7\%)\tnote{*}} & \textbf{251 (0.14\%)\tnote{*}} &\textbf{43 (0.23\%)\tnote{*}} & \textbf{10 (3.60\%)\tnote{*}} \\\hline

6 & SR && X & X & 113 & 39 & 16 & 8 \\\hline
7 & ESR & X & X & X & \textbf{92 (3.62\%)\tnote{**}} & \textbf{25 (0.45\%)\tnote{**}} & \textbf{12 (1.73\%)\tnote{**}} & \textbf{6 (14.63\%)\tnote{**}} \\\bottomrule

\end{tabular}
\begin{tablenotes}
       \item [*] as per Eq. 1
       \item [**] as per Eq. 2
     \end{tablenotes}
\label{tab:search2}
\end{threeparttable}
\end{table*}

\subsection{Temporal Trends}

\begin{figure}
\begin{centering}
\centerline{\includegraphics[width=\linewidth]{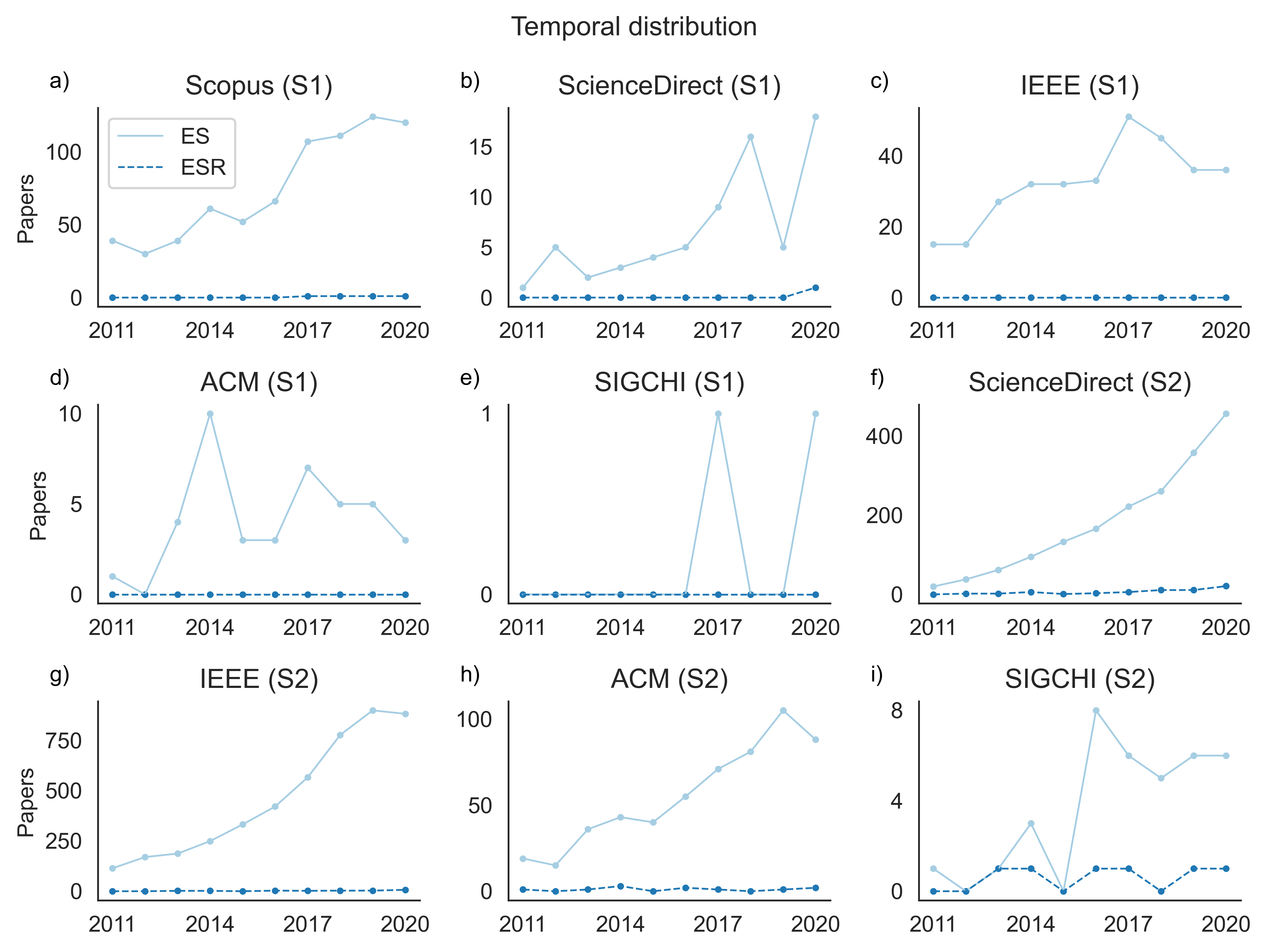}}
\caption{Temporal trends in smart home energy efficiency research (ES) and corresponding engagement with rebound effects (ESR) for Searches 1 (a-e) and 2 (f-i). Although the searches were conducted on all articles indexed in the respective databases, we emphasized the last ten years to examine recent trends in this growing field.}
\label{fig:search_1_and_2}
\Description{Nine graphs that all present the yearly development of the intersections ES and ESR in the 5 databases of Search 1 and the 4 databases of Search 2, respectively. The graphs cover the 10-year timespan 2011-2020. The vast majority of the graphs show a monotonously increasing curve for the ES intersection, but a curve that remains flat (and very close to zero) for the intersection ESR.}
\end{centering}
\end{figure}

As smart homes are an important yet relatively new area of research, it is possible that rebound effects have started to receive attention only in the last couple of years and that we are on a path towards more holistically sustainable research. However, Figure \ref{fig:search_1_and_2} shows that this is also not the case. Although our selected databases depict a net increasing trend in smart home energy efficiency research, most papers do not mention rebound effects. It is evident from Figure~\ref{fig:search_1_and_2} that, since 2011, smart home research has not actively engaged with rebound effects, i.e., yielding zero papers annually across large databases; whereas in other cases, the trend in engagement is stagnant. Scopus, for instance, yielded a single paper for each year from 2017 to 2020 in \textit{Search 1}: clearly, the engagement is not increasing with time. Albeit on quite a low overall level, SIGCHI displays a more consistent (if stagnant as well) rebound engagement. This highlights a critical gap, which is also consistent across \textit{Search 2}. In particular, \textit{Search 2} delineates a significant growth in smart home energy efficiency research in the last ten years; however, such growth is not evident in the ESR intersection. In addition, SIGCHI yielded a greater proportion of research mentioning rebound effects, yet we could not find any signs of growth in attention paid to rebound effects. 

\subsection{Rebound Effects in Computing Literature: The SIGCHI Library}

The SIGCHI library represents a clear outlier compared to the other databases: from the (relatively few) 278 energy efficiency papers in \textit{Search 2}, 10 (3.60\%) address rebound effects. This trend seems even stronger for SIGCHI's energy efficiency research in smart homes: From the 41 papers addressing this field, 6 (14.63\%) consider rebound effects---a vastly larger percentage than the 0.45\% in IEEE and 1.73\% in ACM overall. This raises the question how exactly the papers returned in \textit{Search 2} for this intersection consider rebound effects. Table~\ref{tab:HCIpapers} summarises the papers from our SIGCHI library that contain the smart home, energy efficiency and rebound effect keywords (ESR intersection) at some point in their full texts\footnote{The table lists 5 instead of 6 papers, as one of the results turned out to be the full proceedings of CSCW'17, which contain the keywords, but spread across different papers.}. A closer reading reveals that these papers refer to rebound effects but typically do not place them at the core of the analysis. For example, a recent study of smart meters in use in households discusses energy savings at length but only mentions rebound effects in passing: Sometimes ``users wanted to heat the house a bit more than they would do normally in order to benefit from low prices (typically termed the rebound effect)''~\cite{alan2016too}. Pricing was much more on participants' minds than efficiency. Another study posed an open question about rebound effects as an afterthought: ``How to deal with rebound effects, to ensure efficiency gains are not used in additional emissions elsewhere?''~\cite{ahlers2020challenges}. 

\begin{table*}
\centering
\begin{threeparttable}
\renewcommand{\arraystretch}{1.5}
\caption{SIGCHI papers that contain the smart home, energy efficiency and rebound effect keywords in their full text}
\begin{tabular}{cp{1.3cm}p{10.9cm}}

\toprule
\textbf{Year} & \textbf{Citation} & \textbf{Summary description}\\\midrule
2013 & 
\cite{rodden2013home} & In this study, the authors explore UK consumers' attitudes towards future smart energy infrastructures that combine smart meters with software agents. Use a demonstration and animated sketches, they provide design guidelines to address an identified distrust of energy companies. The rebound effect is briefly mentioned within the motivation of the research. \\\hline
2014 &
\cite{pargman2014rethinking} & This wide-ranging discussion of sustainability in HCI contrasts a set of conceptual frameworks for sustainability to reorient sustainable HCI on a new, more ecological foundation. A well-referenced discussion of rebound effects is part of a broader critique of sustainable HCI research that draws narrow boundaries around its object of interest. \\\hline
2016 &
\cite{alan2016too} & This empirical study observed over the duration of a month how residents used smart thermostats to control the heating in their homes, comparing the effects of three design variations to understand the implications of a set of design choices. The rebound effect is mentioned in passing. \\\hline
2019 & \cite{stevens2019using}
& This co-design study investigates how people use their travel and activity time, as well as their visions for using time in driverless cars. As part of the study the participants were encouraged to engage with interior designs for these cars. The rebound effect is mentioned in the context of value of time theories.  \\\hline
2020 &
\cite{ahlers2020challenges} & A position paper that discusses open questions and existing challenges around urban mobility integration, with a focus on sustainability and emission reductions. Rebound effects are listed as part of these open questions/challenges.  \\\bottomrule
\end{tabular}

\label{tab:HCIpapers}
\end{threeparttable}
\end{table*}

\subsection{Rebound Effects outside of Computing Literature}

What can we learn from the small number of papers that \emph{focus} on rebound effects in a smart home energy efficiency context? We first look at the papers from the general literature outside of computing which \textit{centrally} address rebound effects. These papers, listed in Table~\ref{tab:papers}, have all been published since 2017. Together with the insight that they aim to answer fundamental research questions (e.g., ``Does the rebound effect exist in smart homes and what is the size of the rebound effect?''~\cite{rau_influence_2018}), it becomes apparent that the topic has only recently started to receive attention and that many important, especially more nuanced questions have not yet been addressed. The computing community in particular does not seem to have this on its agenda, as none of the papers in Table \ref{tab:papers} was published at a computing venue. The more slippery indirect and structural rebound effects (see Section~\ref{sec:rebound-indirect}), however, and their often subtle and intricate consequences (both environmental and societal) remain largely unaddressed across research fields; Section~\ref{sec:disc-eff-suff} touches on this topic.

\begin{table*}
\centering
\begin{threeparttable}[b]
\renewcommand{\arraystretch}{1.5}
\caption{Scopus papers containing the smart home, energy efficiency and rebound effect keywords in title, abstract or author keywords}
\begin{tabular}{cp{1.3cm}p{10.9cm}}

\toprule
\textbf{Year} & \textbf{Citation} & \textbf{Summary description}\\\midrule
2017 & % Kumar, T., \& Mani, M. (2017). An energy-neutrality based evaluation into the effectiveness of occupancy sensors in buildings: an integrated life-cycle study. In Vol. II, PLEA Conference Proceedings: Design to Thrive. Edinburgh: PLEA (pp. 2579-2586).
\cite{kumar_energy-neutrality_2017} & The authors conducted a life-cycle assessment to evaluate occupancy sensors for lighting loads in office buildings; based on case studies they also examined the role of designers in this context. Guided by their findings and relevant literature, the authors propose a model framework for smart homes and their components. \\\hline
2018 & \cite{rau_influence_2018} % Chen, K. J., Li, Z., Lu, T. P., Rau, P. L. P., \& Huang, D. (2018, July). Influence of Rebound Effect on Energy Saving in Smart Homes. In International Conference on Cross-Cultural Design (pp. 266-274). Springer, Cham. 
& This study performed experiments with student participants using software simulations of smart homes to better understand the occurrence of rebound effects in smart homes. Based on the obtained results, the authors offer recommendations for rebound effect mitigation. \\\hline
2019 & % Beltram, L., Christensen, M. H., \& Li, R. (2019, November). Demonstration of heating demand peak shaving in smart homes. In Journal of Physics: Conference Series (Vol. 1343, No. 1, p. 012055). IOP Publishing.
\cite{beltram_demonstration_2019} & The paper outlines the results from field tests in which heating algorithms were employed in Danish apartments for flexible demand. User feedback was gathered and the algorithms were refined during the study, leading to better system performance. \\\hline
2020 & % Walzberg, J., Dandres, T., Merveille, N., Cheriet, M., \& Samson, R. (2020). Should we fear the rebound effect in smart homes?. Renewable and Sustainable Energy Reviews, 125, 109798.
\cite{walzberg_should_2020} & In this study, agent-based model simulations were run to assess the rebound effect in smart homes, with emphasis on the impact of contextual aspects of electricity (e.g., cost, demand). The results show that such variables can significantly change the size of occurring rebound effects. \\\bottomrule
\end{tabular}

\label{tab:papers}
\end{threeparttable}
\end{table*}

The papers from Table~\ref{tab:papers} point to some interesting behavioural dynamics, effect sizes and gaps in the literature, some of them of potentially high relevance for HCI. Walzberg et al.~\cite{walzberg_should_2020}, for example, conducted an in-depth analysis of rebound effects, including the distinction between direct and (especially causal factors and dynamics of) indirect effects in smart homes, and the importance of suitable metrics for smart electricity management. The authors suggest that rebound effects do not ``cancel out all the benefits of energy efficiency''; in other words, that smart home efficiency gains do not result in backfire. They advise and promote nevertheless the consideration of factors that decrease rebound effects in smart home design, especially related to user behaviours that tend to trigger them during certain periods; considerations that seem highly relevant for the HCI community. In another study that investigates rebound effects in smart homes, student participants were invited to use smart home software simulations~\cite{rau_influence_2018}. For both of their experimental setups, the results show significant rebound effects. However, as they were using simulations rather than energy data gathered in a real-world smart home context, the ecological validity of this study is low, a fact that the authors seem unaware of; follow-up work is needed to confirm the findings outside of a lab setting. The only study in Table~\ref{tab:papers} that was carried out with occupants in the context of their everyday life is the one by Beltram et al.~\cite{beltram_demonstration_2019}. While their iteratively improved algorithms led to better system performance, no experimental conclusions, especially in relation to the rebound effect, were drawn.

\section{Discussion} \label{sec:discuss}

Investigating the prevalence of rebound effect considerations in energy efficiency research in general and in smart home energy efficiency research in particular, our study consisted of two searches, covering engineering publications as reflected in the IEEE and ACM databases, HCI-specific publications, and general scientific literature as covered by Scopus and ScienceDirect. The searches show that rebound effects are poorly represented in smart home energy efficiency research. While smart home research flourished, attention to rebound effects drifted out of view until recently. Renewed attention to their importance is now providing important impetus and starting points that offer promising opportunities and directions for HCI research.

\subsection{Unpacking Rebound Effects}

To support HCI research, we offer here a taxonomy that synthesises findings from our literature mapping and the wider literature to provide concrete starting points for the HCI community to engage with and further the topic of rebound effects. We envision this taxonomy as the beginning of a much-needed, systematic stream of research to develop guidelines and conceptual frameworks to account for rebound effects both reactively and prospectively:

\begin{enumerate}
\item A posteriori, we need frameworks and empirical methods to measure the size of rebound effects in real cases. Such measurements can inform the (un)design~\cite{pierce2012undesigning} of existing technology, including its upscaling/downscaling, removal, replacement and restoration.
\item A priori, we need tools to evaluate the \textit{potential} for rebound effects in given circumstances. Mitigation strategies are essential at all times, but in particular before technology design and development. In some cases, such as those in which rebound effects are expected to cancel out efficiency savings, the implication could even be not to design technology at all but to focus on low-tech or no-tech solutions~\cite{baumer2011implication}.
\end{enumerate}

\begin{table*}
\centering
\caption{A taxonomy for addressing rebound effects in HCI}
\footnotesize
\begin{tabularx}{\linewidth}{cXXX}
\toprule
\textbf{Action} & \textbf{Direct} & \textbf{Indirect} & \textbf{Structural} \\\midrule
Identify & 
It is paramount to identify in which cases/contexts efficiency savings ultimately increase the consumption of a particular good. To add to the often technical literature, in-situ~\cite{carroll2013wild} and living lab research~\cite{buhl2017rebound} would allow HCI researchers to study the outcome of strategies and interventions that might trigger rebound effects (e.g., behaviour change~\cite{turner2013rebound}, automation~\cite{alan2016too}).
&
Indirect rebound occurs when (e.g., efficiency) savings are spent on other, potentially more resource-intensive goods and services, a dilemma HCI researchers have started to recognise~\cite{widdicks2019streaming, remy2018evaluation}. Their empirical work could help uncover resource flows. Indirect rebound is more difficult to account for than direct rebound, so many studies focus only on the latter~\cite{chitnis2015living}.
&
Structural rebound, also called economy-wide or macro-level rebound, affects the socio-economic system as whole; it shows the impact of individuals' behaviour on a structural level~\cite{hilty2015ict}. HCI researchers have been called to recognise structural rebound as a potential outcome of their designs~\cite{remy2018evaluation,pargman2014:rebound} and to engage with systemic impacts more generally~\cite{nathan2008envisioning}.\\\midrule

Measure 
&
Direct rebound effects are commonly estimated through simulations and econometric research, including general equilibrium models~\cite{azevedo2014consumer}. Carefully designed quasi-experimental approaches offer a promising alternative~\cite{sorrell_jevons_2009} and could help address the lack of empirical work~\cite{turner2013rebound,saunders2000:rebound}. Longitudinal HCI studies could feed into interdisciplinary collaborations.
&
While indirect rebound is more difficult to measure, simulations and econometric approaches still provide valid methodologies~\cite{turner2009negative};
little empirical work exists, but is called for~\cite{turner2013rebound, saunders2000:rebound}.
Estimates for both direct and indirect effects vary vastly: in e.g., households, the combined range spans 5\% - 175\%, reflecting a high degree of uncertainty~\cite{nassen_quantifying_2009, chitnis2015living}.
&
The size of structural rebound is insufficiently researched~\cite{stern2020large,turner2013rebound}, but globally, economic activity is fully coupled to material footprint~\cite{wiedmann_material_2015}. Simulations and econometric research have produced mixed results: estimates range from backfire to super-conservation~\cite{saunders2013historical,turner2013rebound}. Empirical research is called for on the level of firms and economies~\cite{lange2019economy}.
\\\midrule

Explain
&
To understand the mechanisms behind rebound effects and create rigorous theoretical frameworks, it is important to identify relevant factors and how they shape user behaviour~\cite{turner2013rebound}. Potential factors and, thus, interesting starting points for HCI research include energy supply responses~\cite{turner2013rebound}, building types~\cite{guerra2013occupant} and personal values~\cite{seebauer2018psychology}.
&
Indirect rebound can be challenging to explain as it is more fluid: if one outcome variable is constrained, the impetus behind the rebound may find a path to express itself in an unconstrained variable. Identified mechanisms, including the induction effect, income effect and substitution effect~\cite{coroama_digital_2019}, provide pointers for future research, in HCI and beyond.
&
To identify and explain structural rebound, system dynamics constitute a promising approach~\cite{penzenstadler2014safety,penzenstadler2018software}; here, HCI researchers could draw on their experience with systemic methods~\cite{bornes2022could}. Theory already provides some guidance on the factors influencing structural rebound, but it does not impose much constraint on the size of the effects~\cite{stern2020large}. 
\\\midrule

Mitigate
&
To prevent direct rebound from limiting or outweighing efficiency savings, it needs to be mitigated. HCI researchers could apply their design skills to counteract underlying mechanisms by e.g., working with information flows as a key leverage point~\cite{penzenstadler2018software}, encouraging frugal use and raising awareness~\cite{kisselburgh2020hci}. And their prototyping skills enable evaluations without infrastructure cost~\cite{pierce2012beyond}.
&
Preist et al.~\cite{preist2016understanding} provide a lens to better understand both the mechanisms behind wasteful behaviour leading to an increasing ICT footprint, and design principles and strategies to address these mechanisms. These reflections could be extended to the not quite so different mechanisms behind indirect rebound, and induce a shift towards more radically sustainable design practices.
&
Structural rebound is a central reason why sustainability cannot be achieved by decoupling strategies alone: instead of relying on efficiency strategies, sufficiency strategies and policies for deeper structural change are needed~\cite{hilty2011sustainability,t_parrique_decoupling_2019,hickel_is_2020}. An example is the introduction of a global carbon constraint~\cite{freitag2021climate}; in this context, HCI could contribute towards Green Policy Informatics~\cite{bremer2022have}. 
\\\bottomrule
\end{tabularx}
\label{tab:matrix}
\end{table*}

The taxonomy is shown in Table~\ref{tab:matrix} and takes the form of a systematic array with concrete opportunities for HCI to effectively apply their skillset towards the identification, measurement, explanation and mitigation of direct, indirect and structural rebound effects, which are described in section \ref{sec:retypes}. Examples of such opportunities are gaps and calls in the literature, as well as challenges and relevant methodological suggestions by type of rebound. 

Across all types of rebound, Table~\ref{tab:matrix} shows that more empirical research is called for, e.g.,~\cite{turner2013rebound,saunders2000:rebound,lange2019economy}. While the complex and systemic nature of rebound makes this a challenging task, the literature offers concrete tools to tackle it, including but not limited to quasi-experimental approaches~\cite{sorrell_jevons_2009}, living lab research~\cite{buhl2017rebound} and systems thinking~\cite{penzenstadler2018software}. Covering both quantitative and qualitative research skills, these approaches fall under the expertise of many HCI researchers. And the field has, in fact, acknowledged the importance of rebound in its research: even when they are complex and not yet sufficiently understood, HCI researchers have been encouraged to engage with rebound effects when designing, deploying and evaluating ICT systems~\cite{pargman2014rethinking,remy2018evaluation,widdicks2019streaming}. So not only can they provide insights for policy makers and other research fields, but they can improve the outcomes of their own projects as well.

Several studies mentioned in Table~\ref{tab:matrix} point to the potential to better understand the causal factors of rebound and to identify and evaluate \textit{rebound patterns} that can inform research and design. For example, rebound effects are \textit{enabled and shaped} by such features as flexible uses, the potential for new types of uses that permit induced demand, and the savings of resources that can otherwise be redeployed~\cite{coroama_skill_2020}. On the other hand, they are \textit{constrained} by factors such as natural caps or limits on objectives, such as comfortable temperature zones for human living or existing real-estate costs~\cite{coroama_digital_2019}. After all, rebound is not an arbitrary randomizer. In many cases, natural caps exist that place constraints on the amount of rebound we can reasonably expect. Research in energy policy provides conceptual guidance for methods and macro-level views~\cite{saunders_rebound_2012} that can structure and inform HCI research. For example, there is a significant difference between lighting efficiency gains and heating gains in that lighting has additional uses and therefore is subject to induced demand, whereas heating is limited by our temperature comfort zones. 

Digitalisation-triggered rebound occurs both within the ICT sector, and also outside it, in society and economy at large. The inner-sectoral rebound is represented by e.g., the induced material and energy consumption of smart technologies such as those deployed in smart homes. It is also reflected in Koomey's law~\cite{koomey2011:law} that shows that the exponential efficiency gains of computation (as reflected by the well-known Moore's law) have not led to close-to-zero energy consumption for computation, but to an equally exponential growth in the complexity of computations performed---in other words, a direct rebound effect of 100\%. Our taxonomy and the HCI opportunities presented in Table~\ref{tab:matrix} are agnostic to whether the rebound triggered by digitalisation occurs within or outside the ICT sector, although the larger need, and thus opportunities, might present themselves across all other sectors.

\subsection{A Call to Action for the HCI community}

Following a set of critiques papers, there is evidence that the sustainable HCI community has recently moved away from persuasive and efficiency-focused projects (which accounted for the majority of publications in the field in 2010~\cite{disalvo2010mapping}) and further embraced third-wave, qualitative work \cite{bremer2022have}. This work is often more exploratory in nature, with the aim to gather insights rather than to realise pre-determined behavioural or efficiency targets. It also captures the community's understanding that the deployment of technology does not happen in an (easily modelled) vacuum; instead, it is situated within the complicated everyday lives of people who are themselves shaped by cultures, demands and individual differences---and it often is the situatedness that makes energy efficiency measures backfire. Communicating this understanding to policy-makers, other stakeholders, and potentially more technically-oriented fields could prevent at an early stage misleading research outputs in those fields as well as potentially misled policy measures and incentives as a result of such incomplete analyses. After all, our literature mapping has shown that energy efficiency (in smart homes and beyond) is a vast area of research; Scopus returned almost a quarter-million papers for \emph{Search 1} alone. 

In particular, the lack of studies that focus on occupants in the context of their everyday life presents a unique opportunity for HCI research that has commonly been described as research `in the wild'~\cite{rogers2017:wild,carroll2013wild}; due to the term's colonial connotations and associated exploitative practices~\cite{lewis1973anthropology,ssozi2016enough}, we, the authors, are opting to instead describe such research as \emph{in situ}, a term also often deployed in HCI research~\cite{carroll2013wild,consolvo2007conducting}. In-situ HCI research can help other disciplines understand the impact of rebound effects by studying situated behaviours in smart home environments. As established in section~\ref{sec:background}, HCI researchers come equipped with expertise on both efficiency work and the nuanced ways in which socio-technical factors undermine energy savings in smart homes. Taken together with our finding that HCI shows more awareness of rebound effects than all of the other research communities we included in our search, this puts the HCI community in a position where it can make a real difference: if digital technologies do not (consistently) enable the envisioned efficiency savings, this needs to be acknowledged when climate change mitigation strategies are selected.

For various methodological reasons, assessing the indirect effects of digitalisation (both the environmentally beneficial and detrimental ones) is challenging~\cite{coroama-method2020}, and rebound effects are no exception: ``Environmental impacts that arise when technologies co-evolve with everyday practices are not easily predictable. This seems to be one reason why the existing literature [\ldots] contains relatively few or vague recommendations to policy-makers and other stakeholders''~\cite{borjesson_rivera2014}. From the four promising methodological approaches put forward by Börjesson Rivera et  al.~\cite{borjesson_rivera2014}, two are the scenario method and ethnographic methods as well as anthropological theory, respectively. HCI researchers come equipped to apply these methods, arguably better than their colleagues in any other computing discipline.

The many factors that drive rebound effects (and their interplay) add to the complexity of the required research and likely constitute one of the reasons why there exist few empirical studies to date. To deal with such complexity, a shift from computational thinking to system-level thinking has been encouraged among technologists: ``the failure to think systemically is a critical weakness in our understanding of the transformations needed to achieve sustainability''~\cite{easterbrook2014computational}. Besides contextually embedded in-situ research, HCI methods can help to explore and speculate about potential impacts of smart home technologies, including their undesirable consequences, within and beyond the constraints of contemporary technologies and social life (e.g.,~\cite{soden2021we,watson2021hci,ilstedt2014altering}). And they can be used to successfully implement top-down measures (e.g., stricter climate policies as in `Green Policy Informatics'~\cite{bremer2022have}), which will likely have a key role in climate change mitigation.

\subsection{Efficiency, Sufficiency, and Justice}
\label{sec:disc-eff-suff}

Clearly, the ongoing digitisation seems not to have mitigated our environmental issues, as its globalising and accelerating rebound effects probably outweigh the benefits. As Santarius~\cite{santarius2017} notices, ``Humanity's ecological footprint keeps growing although we have already digitalized significant parts of our economy and society over the past years. It seems that digitalization is not relaxing but rather reshaping societal metabolism in a way that tends to rebound on global energy and resource demand: Gains in efficiency are more than outweighed by the increase in consumption due to new digital services or falling prices caused by more efficient production processes''~\cite{santarius2017}. A paper fittingly subtitled \emph{Why Digitalization Will Not Redeem Us Our Environmental Sins} adds: ``Digitalization is unlikely to be the environmental silver bullet it is sometimes claimed to be. On the contrary, the way digitalization changes society, making it ever faster, more connected, and allowing us unprecedented levels of efficiency might in fact lead to a backfire [\ldots] For most manifestations of digitalization, a strong digital rebound seems to be the rule rather than the exception''~\cite{coroama_digital_2019}.

Together with other factors, rebound effects obstruct the hopes that digitalisation and other technical innovations can `dematerialise' economic activity from resource consumption so that the economy can keep growing while its material footprint remains constant or shrinks. Other factors include the systematically underestimated impact of services, the limited potential of recycling, as well as insufficient and
inappropriate technological change~\cite{t_parrique_decoupling_2019}. Rebound effects are one of these central factors standing against the feasibility of absolute decoupling, and they require ``an in-depth and systemic consideration and anticipation [...] in the design of sustainability policies''~\cite[p.~40]{t_parrique_decoupling_2019}. Unfortunately, according to all available evidence, ``it is safe to say that the type of decoupling acclaimed by green growth advocates is essentially a statistical figment'' \cite[p. 31]{t_parrique_decoupling_2019}, \cite{haberl_systematic_2020,vaden_decoupling_2020,wiedmann_material_2015,hickel_is_2020,hickel_national_2022,fanning_social_2022}.  Together, these insights reinforce the message that a more profound shift is needed to truly advance toward sustainable societies \cite{mann_shifting_2018,knowles_cyber-sustainabilitytowards_2013, pargman2014rethinking, hickel_less_2020, nardi_design_2019} and confirm how important it is to recognise and evaluate the effects of technology-focused interventions through a systemic view.

In this context, the combination of efficiency and sufficiency strategies could form the basis of the profound shift that is called for: When technical efficiency concurs with limited input resources, real progress could be made towards long-term sustainable resource usage~\cite{hilty2011sustainability}. Digital technologies could then be used to replace or improve existing activities without spurring additional activities. In fact, Freitag et al.~\cite{freitag2021climate} argue that ``[i]f a global carbon constraint was introduced, we could be certain that rebound effects would not occur, meaning that productivity improvements through ICT-enabled efficiencies both within the ICT sector and the wider economy would be realised without a carbon cost''. Such a nondiscriminatory approach that simply imposes global carbon constraints, however, can perpetuate or even intensify existing disparities, as under-represented or marginalized communities would be disproportionately affected~\cite{sovacool2023:pluralizing-justice}. 
Well-meant measures to mitigate this risk can themselves have paradoxical counter-effects in turn, such as anti-racist campaigns unwillingly reinforcing the very disparities they set out to address~\cite{lennon2021:energy-justice} or even the concept of energy justice itself, which often ``fails to adequately account for intersecting dimensions of power and inequality, such as gender, race, class, Indigeneity, ethnicity, sexuality, ability status, colonial history, and caste, among other statuses within the world system''~\cite{sovacool2023:pluralizing-justice}. 
Rebound effects, and in particular the elusive indirect rebound with potential ramifications throughout the entire society, could be particularly prone to amplify existing disparities. 
The call for sufficiency and effective climate policies is not novel in the computing communities, e.g. ~\cite{hilty2011sustainability,bremer2022have}.
Our call for HCI, however, is to shine light on the critically important topic of rebound effects from a systems thinking perspective~\cite{widdicks2023:rebound-systems}, but not one focused exclusively on environmental matters, but one with justice considerations at its very core.

\subsection{Limitations}

Like any systematic mapping study, the scope of ours by necessity exhibits some limitations. The databases do not cover all existing literature on smart homes, energy efficiency and rebound effects. This means that we may have missed some of the published work in these areas. Second, it is unclear to what extent the result sets from the different databases for each search overlap (with the exception of the SIGCHI library, an ACM subset by design). Lastly, due to the reliance of specific keywords as indicators for both awareness and research focus, it is likely that our result sets are affected by a small number of false positives (i.e., papers that include specific keywords, but do not engage in analysis of the subject) and false negatives (i.e., papers that cover the subject area but do not use the search terminology, for example, papers covering energy efficiency and consumption without using the term ‘energy efficient’). For example, one of the 6 SIGCHI results in the ESR set of \emph{Search 2} turned out to be the full proceedings of CSCW'17, which contain all keywords spread across different papers. Nonetheless, the covered databases are among the most comprehensive sources available, and combined, the number of missed papers should be minimal. In addition, full-text search and manual verification address false positives, the terminology of rebound effects is well established, and the relative lack of attention is widespread and consistent across all searches. The trends we identified are unlikely to be affected by minor shifts in time or coverage. Therefore, these limitations do not threaten the validity of the findings.

\section{Conclusion}

As a long-standing concept in the economics literature, rebound effects capture the impact of socio-technical factors that lead to reduced energy savings after the introduction of energy efficiency measures. With the perception of digital technologies as a key enabler of these savings, considering rebound effects in efficiency work is crucial. In our paper, we present the results from a literature mapping that show minimal considerations of rebound effects in scientific and computing databases within the context of smart home energy efficiency work; of the included research communities, it is HCI that appears to be most aware of rebound effects. In its recent history, the community has already produced an impressive set of qualitative work that captures how smart homes can undermine energy savings through new lifestyle expectations and changing behaviours, as well as a large set of persuasive and efficiency-focused studies. To help quantify the impact of the factors identified in HCI's qualitative work, move beyond simple proxies in efficiency-focused studies and have a language to share findings with other disciplines, we call the community to actively adopt the concept of rebound effects for our research. The taxonomy and discussion in Table~\ref{tab:matrix} provide a starting point.

By doing so, we believe that HCI researchers are uniquely equipped to bridge the existing awareness gap to help other disciplines avoid energy and sociotechnical rebounds. As an important complement to qualitative insights in decision-making processes (for e.g., policy-makers), HCI researchers can also support colleagues in these disciplines to establish context-specific, accurate rebound quantifications (which can provide a basis for finding patterns and averages, making predictions, testing causal relationships and generalising results to wider populations). They can use their tools and methods, including empirical methods for smart home research and design tools to capture and visualise complex estimates, to shape a more focused, nuanced understanding of energy savings that are real, not offset by rebound. In other words, they can help establish how viable energy savings (in smart homes and beyond) are, as a significant contribution towards effective climate change mitigation strategies.

\begin{acks}
Christina Bremer and Christoph Becker gratefully acknowledge the support of their respective funders, the Leverhulme Centre for Material Social Futures Research (DS-2017-036) and the Natural Sciences and Engineering Research Council of Canada (RGPIN-2016-06640). We, the authors, also thank the reviewers for their thoughtful feedback that has enabled us to improve an earlier version of the manuscript, Anjana Mohandas Sheeladevi for contributions to initial conversations, and Bran Knowles, Adrian Friday, Steve Easterbrook and Matthias Finkbeiner for offering research guidance.

CRediT authorship contribution statement Christina Bremer: Conceptualization, Methodology, Validation, Formal analysis, Investigation, Writing - Original Draft, Writing - Review \& Editing, Visualization, Supervision, Project administration. Harshit Gujral: Conceptualization, Methodology, Software, Formal analysis, Investigation, Writing - Original Draft, Writing - Review \& Editing, Visualization. Michelle Lin: Conceptualization, Writing - Original Draft. Lily Hinkers: Investigation. Christoph Becker: Conceptualization, Methodology, Formal analysis, Investigation, Writing - Original Draft, Writing - Review \& Editing, Supervision. Vlad C. Coroamă: Conceptualization, Methodology, Formal analysis, Investigation, Writing - Original Draft, Writing - Review \& Editing, Supervision. 
\end{acks}

%% The next two lines define the bibliography style to be used, and the bibliography file.
\bibliographystyle{ACM-Reference-Format}
\bibliography{bibliography}


\begin{thebibliography}{124}

%%% the Web address.

\ifx \showCODEN    \undefined \def \showCODEN     #1{\unskip}     \fi
\ifx \showDOI      \undefined \def \showDOI       #1{#1}\fi
\ifx \showISBNx    \undefined \def \showISBNx     #1{\unskip}     \fi
\ifx \showISBNxiii \undefined \def \showISBNxiii  #1{\unskip}     \fi
\ifx \showISSN     \undefined \def \showISSN      #1{\unskip}     \fi
\ifx \showLCCN     \undefined \def \showLCCN      #1{\unskip}     \fi
\ifx \shownote     \undefined \def \shownote      #1{#1}          \fi
\ifx \showarticletitle \undefined \def \showarticletitle #1{#1}   \fi
\ifx \showURL      \undefined \def \showURL       {\relax}        \fi
% invisible to TeX
\providecommand\bibfield[2]{#2}
\providecommand\bibinfo[2]{#2}
\providecommand\natexlab[1]{#1}
\providecommand\showeprint[2][]{arXiv:#2}

\bibitem[A.~Greening et~al\mbox{.}(2000)]%
        {a_greening_energy_2000}
\bibfield{author}{\bibinfo{person}{Lorna A.~Greening}, \bibinfo{person}{David~L. Greene}, {and} \bibinfo{person}{Carmen Difiglio}.} \bibinfo{year}{2000}\natexlab{}.
\newblock \showarticletitle{Energy efficiency and consumption -— the rebound effect -— a survey}.
\newblock \bibinfo{journal}{\emph{Energy Policy}} \bibinfo{volume}{28}, \bibinfo{number}{6} (\bibinfo{date}{June} \bibinfo{year}{2000}), \bibinfo{pages}{389--401}.
\newblock
\showISSN{0301-4215}
\urldef\tempurl%
\url{https://doi.org/10.1016/S0301-4215(00)00021-5}
\showDOI{\tempurl}


\bibitem[Ahlers(2020)]%
        {ahlers2020challenges}
\bibfield{author}{\bibinfo{person}{Dirk Ahlers}.} \bibinfo{year}{2020}\natexlab{}.
\newblock \showarticletitle{Challenges of Sustainable Urban Mobility Integration}. In \bibinfo{booktitle}{\emph{22nd International Conference on Human-Computer Interaction with Mobile Devices and Services}}. \bibinfo{pages}{1--3}.
\newblock


\bibitem[Alan et~al\mbox{.}(2016)]%
        {alan2016too}
\bibfield{author}{\bibinfo{person}{Alper~T Alan}, \bibinfo{person}{Mike Shann}, \bibinfo{person}{Enrico Costanza}, \bibinfo{person}{Sarvapali~D Ramchurn}, {and} \bibinfo{person}{Sven Seuken}.} \bibinfo{year}{2016}\natexlab{}.
\newblock \showarticletitle{It is too hot: An in-situ study of three designs for heating}. In \bibinfo{booktitle}{\emph{Proceedings of the 2016 CHI Conference on Human Factors in Computing Systems}}. \bibinfo{pages}{5262--5273}.
\newblock


\bibitem[Alcott(2005)]%
        {alcott2005:jevons}
\bibfield{author}{\bibinfo{person}{Blake Alcott}.} \bibinfo{year}{2005}\natexlab{}.
\newblock \showarticletitle{Jevons' paradox}.
\newblock \bibinfo{journal}{\emph{Ecological Economics}}  \bibinfo{volume}{54} (\bibinfo{year}{2005}), \bibinfo{pages}{9--21}.
\newblock
\urldef\tempurl%
\url{https://www.sciencedirect.com/science/article/pii/S0921800905001084}
\showURL{%
\tempurl}


\bibitem[Azevedo(2014)]%
        {azevedo2014consumer}
\bibfield{author}{\bibinfo{person}{In{\^e}s~ML Azevedo}.} \bibinfo{year}{2014}\natexlab{}.
\newblock \showarticletitle{Consumer end-use energy efficiency and rebound effects}.
\newblock \bibinfo{journal}{\emph{Annual Review of Environment and Resources}} \bibinfo{volume}{39}, \bibinfo{number}{1} (\bibinfo{year}{2014}), \bibinfo{pages}{393--418}.
\newblock


\bibitem[Barker et~al\mbox{.}(2009)]%
        {barker2009:macroeconomic-rebound}
\bibfield{author}{\bibinfo{person}{Terry Barker}, \bibinfo{person}{Athanasios Dagoumas}, {and} \bibinfo{person}{Jonathan Rubin}.} \bibinfo{year}{2009}\natexlab{}.
\newblock \showarticletitle{The macroeconomic rebound effect and the world economy}.
\newblock \bibinfo{journal}{\emph{Energy Efficiency}} \bibinfo{volume}{2}, \bibinfo{number}{4} (\bibinfo{date}{May} \bibinfo{year}{2009}), \bibinfo{pages}{411--427}.
\newblock
\showISSN{1570-6478}
\urldef\tempurl%
\url{https://doi.org/10.1007/s12053-009-9053-y}
\showDOI{\tempurl}


\bibitem[Baumer and Silberman(2011)]%
        {baumer2011implication}
\bibfield{author}{\bibinfo{person}{Eric~PS Baumer} {and} \bibinfo{person}{M~Six Silberman}.} \bibinfo{year}{2011}\natexlab{}.
\newblock \showarticletitle{When the implication is not to design (technology)}. In \bibinfo{booktitle}{\emph{Proceedings of the 2011 CHI Conference on Human Factors in Computing Systems}}. \bibinfo{pages}{2271--2274}.
\newblock


\bibitem[Becker et~al\mbox{.}(2018)]%
        {becker_estimating_2018}
\bibfield{author}{\bibinfo{person}{Vincent Becker}, \bibinfo{person}{Wilhelm Kleiminger}, \bibinfo{person}{Vlad~C. Coroamă}, {and} \bibinfo{person}{Friedemann Mattern}.} \bibinfo{year}{2018}\natexlab{}.
\newblock \showarticletitle{Estimating the savings potential of occupancy-based heating strategies}.
\newblock \bibinfo{journal}{\emph{Energy Informatics}} \bibinfo{volume}{1}, \bibinfo{number}{S1} (\bibinfo{date}{Oct.} \bibinfo{year}{2018}), \bibinfo{pages}{52}.
\newblock
\showISSN{2520-8942}
\urldef\tempurl%
\url{https://doi.org/10.1186/s42162-018-0022-6}
\showDOI{\tempurl}


\bibitem[Beltram et~al\mbox{.}(2019)]%
        {beltram_demonstration_2019}
\bibfield{author}{\bibinfo{person}{Lucas Beltram}, \bibinfo{person}{Morten~Herget Christensen}, {and} \bibinfo{person}{Rongling Li}.} \bibinfo{year}{2019}\natexlab{}.
\newblock \showarticletitle{Demonstration of heating demand peak shaving in smart homes}.
\newblock \bibinfo{journal}{\emph{Journal of Physics: Conference Series}} \bibinfo{volume}{1343}, \bibinfo{number}{1} (\bibinfo{date}{Nov.} \bibinfo{year}{2019}), \bibinfo{pages}{012055}.
\newblock
\showISSN{1742-6588, 1742-6596}
\urldef\tempurl%
\url{https://doi.org/10.1088/1742-6596/1343/1/012055}
\showDOI{\tempurl}


\bibitem[Berkhout et~al\mbox{.}(2000)]%
        {berkhout_defining_2000}
\bibfield{author}{\bibinfo{person}{Peter H.~G. Berkhout}, \bibinfo{person}{Jos~C. Muskens}, {and} \bibinfo{person}{Jan W.~Velthuijsen}.} \bibinfo{year}{2000}\natexlab{}.
\newblock \showarticletitle{Defining the rebound effect}.
\newblock \bibinfo{journal}{\emph{Energy Policy}} \bibinfo{volume}{28}, \bibinfo{number}{6} (\bibinfo{date}{June} \bibinfo{year}{2000}), \bibinfo{pages}{425--432}.
\newblock
\showISSN{0301-4215}
\urldef\tempurl%
\url{https://doi.org/10.1016/S0301-4215(00)00022-7}
\showDOI{\tempurl}


\bibitem[Binswanger(2001)]%
        {binswanger_technological_2001}
\bibfield{author}{\bibinfo{person}{Mathias Binswanger}.} \bibinfo{year}{2001}\natexlab{}.
\newblock \showarticletitle{Technological progress and sustainable development: what about the rebound effect?}
\newblock \bibinfo{journal}{\emph{Ecological Economics}} \bibinfo{volume}{36}, \bibinfo{number}{1} (\bibinfo{date}{Jan.} \bibinfo{year}{2001}), \bibinfo{pages}{119--132}.
\newblock
\showISSN{09218009}
\urldef\tempurl%
\url{https://doi.org/10.1016/S0921-8009(00)00214-7}
\showDOI{\tempurl}


\bibitem[Bornes et~al\mbox{.}(2022)]%
        {bornes2022could}
\bibfield{author}{\bibinfo{person}{Laetitia Bornes}, \bibinfo{person}{Catherine Letondal}, {and} \bibinfo{person}{Rob Vingerhoeds}.} \bibinfo{year}{2022}\natexlab{}.
\newblock \showarticletitle{Could Systemic Design Methods Support Sustainable Design Of Interactive Systems?}. In \bibinfo{booktitle}{\emph{Relating Systems Thinking and Design 2021 Symposium}}.
\newblock


\bibitem[Bremer et~al\mbox{.}(2022)]%
        {bremer2022have}
\bibfield{author}{\bibinfo{person}{Christina Bremer}, \bibinfo{person}{Bran Knowles}, {and} \bibinfo{person}{Adrian Friday}.} \bibinfo{year}{2022}\natexlab{}.
\newblock \showarticletitle{Have We Taken On Too Much?: A Critical Review of the Sustainable HCI Landscape}. In \bibinfo{booktitle}{\emph{CHI Conference on Human Factors in Computing Systems}}. \bibinfo{pages}{1--11}.
\newblock


\bibitem[Brookes(1990)]%
        {brookes1990:rebound-ghg}
\bibfield{author}{\bibinfo{person}{Len Brookes}.} \bibinfo{year}{1990}\natexlab{}.
\newblock \showarticletitle{The greenhouse effect: the fallacies in the energy efficiency solution}.
\newblock \bibinfo{journal}{\emph{Energy Policy}} \bibinfo{volume}{18}, \bibinfo{number}{2} (\bibinfo{year}{1990}), \bibinfo{pages}{199--201}.
\newblock
\showISSN{0301-4215}
\urldef\tempurl%
\url{https://doi.org/10.1016/0301-4215(90)90145-T}
\showDOI{\tempurl}


\bibitem[Buhl et~al\mbox{.}(2017)]%
        {buhl2017rebound}
\bibfield{author}{\bibinfo{person}{Johannes Buhl}, \bibinfo{person}{Justus von Geibler}, \bibinfo{person}{Laura Echternacht}, {and} \bibinfo{person}{Moritz Linder}.} \bibinfo{year}{2017}\natexlab{}.
\newblock \showarticletitle{Rebound effects in Living Labs: Opportunities for monitoring and mitigating re-spending and time use effects in user integrated innovation design}.
\newblock \bibinfo{journal}{\emph{Journal of cleaner production}}  \bibinfo{volume}{151} (\bibinfo{year}{2017}), \bibinfo{pages}{592--602}.
\newblock


\bibitem[Börjesson~Rivera et~al\mbox{.}(2014)]%
        {borjesson_rivera2014}
\bibfield{author}{\bibinfo{person}{Miriam Börjesson~Rivera}, \bibinfo{person}{Cecilia Håkansson}, \bibinfo{person}{Åsa Svenfelt}, {and} \bibinfo{person}{Göran Finnveden}.} \bibinfo{year}{2014}\natexlab{}.
\newblock \showarticletitle{Including second order effects in environmental assessments of {ICT}}.
\newblock \bibinfo{journal}{\emph{Environmental Modelling \& Software}}  \bibinfo{volume}{56} (\bibinfo{date}{June} \bibinfo{year}{2014}), \bibinfo{pages}{105--115}.
\newblock
\showISSN{13648152}
\urldef\tempurl%
\url{https://doi.org/10.1016/j.envsoft.2014.02.005}
\showDOI{\tempurl}


\bibitem[Carroll and Rosson(2013)]%
        {carroll2013wild}
\bibfield{author}{\bibinfo{person}{John~M Carroll} {and} \bibinfo{person}{Mary~Beth Rosson}.} \bibinfo{year}{2013}\natexlab{}.
\newblock \showarticletitle{Wild at home: The neighborhood as a living laboratory for HCI}.
\newblock \bibinfo{journal}{\emph{ACM Transactions on Computer-Human Interaction (TOCHI)}} \bibinfo{volume}{20}, \bibinfo{number}{3} (\bibinfo{year}{2013}), \bibinfo{pages}{1--28}.
\newblock


\bibitem[Chen et~al\mbox{.}(2018)]%
        {rau_influence_2018}
\bibfield{author}{\bibinfo{person}{Ko-jung Chen}, \bibinfo{person}{Ziyang Li}, \bibinfo{person}{Ta-Ping Lu}, \bibinfo{person}{Pei-Luen~Patrick Rau}, {and} \bibinfo{person}{Dinglong Huang}.} \bibinfo{year}{2018}\natexlab{}.
\newblock \showarticletitle{Influence of {Rebound} {Effect} on {Energy} {Saving} in {Smart} {Homes}}.
\newblock In \bibinfo{booktitle}{\emph{Cross-{Cultural} {Design}. {Applications} in {Cultural} {Heritage}, {Creativity} and {Social} {Development}}}, \bibfield{editor}{\bibinfo{person}{Pei-Luen~Patrick Rau}} (Ed.). Vol.~\bibinfo{volume}{10912}. \bibinfo{publisher}{Springer International Publishing}, \bibinfo{address}{Cham}, \bibinfo{pages}{266--274}.
\newblock
\showISBNx{978-3-319-92251-5 978-3-319-92252-2}
\urldef\tempurl%
\url{https://doi.org/10.1007/978-3-319-92252-2_21}
\showDOI{\tempurl}
\newblock
\shownote{Series Title: Lecture Notes in Computer Science}.


\bibitem[Chitnis and Sorrell(2015)]%
        {chitnis2015living}
\bibfield{author}{\bibinfo{person}{Mona Chitnis} {and} \bibinfo{person}{Steve Sorrell}.} \bibinfo{year}{2015}\natexlab{}.
\newblock \showarticletitle{Living up to expectations: Estimating direct and indirect rebound effects for UK households}.
\newblock \bibinfo{journal}{\emph{Energy Economics}}  \bibinfo{volume}{52} (\bibinfo{year}{2015}), \bibinfo{pages}{S100--S116}.
\newblock


\bibitem[Consolvo et~al\mbox{.}(2007)]%
        {consolvo2007conducting}
\bibfield{author}{\bibinfo{person}{Sunny Consolvo}, \bibinfo{person}{Beverly Harrison}, \bibinfo{person}{Ian Smith}, \bibinfo{person}{Mike~Y Chen}, \bibinfo{person}{Katherine Everitt}, \bibinfo{person}{Jon Froehlich}, {and} \bibinfo{person}{James~A Landay}.} \bibinfo{year}{2007}\natexlab{}.
\newblock \showarticletitle{Conducting in situ evaluations for and with ubiquitous computing technologies}.
\newblock \bibinfo{journal}{\emph{International Journal of Human-Computer Interaction}} \bibinfo{volume}{22}, \bibinfo{number}{1-2} (\bibinfo{year}{2007}), \bibinfo{pages}{103--118}.
\newblock


\bibitem[Coroam\u{a} et~al\mbox{.}(2020)]%
        {coroama-method2020}
\bibfield{author}{\bibinfo{person}{Vlad~C. Coroam\u{a}}, \bibinfo{person}{Pernilla Bergmark}, \bibinfo{person}{Mattias H\"{o}jer}, {and} \bibinfo{person}{Jens Malmodin}.} \bibinfo{year}{2020}\natexlab{}.
\newblock \showarticletitle{A Methodology for Assessing the Environmental Effects Induced by ICT Services: Part I: Single Services} \emph{(\bibinfo{series}{ICT4S2020})}. \bibinfo{publisher}{Association for Computing Machinery}, \bibinfo{address}{New York, NY, USA}, \bibinfo{pages}{36--–45}.
\newblock
\showISBNx{9781450375955}
\urldef\tempurl%
\url{https://doi.org/10.1145/3401335.3401716}
\showDOI{\tempurl}


\bibitem[Coroamă and Mattern(2019)]%
        {coroama_digital_2019}
\bibfield{author}{\bibinfo{person}{Vlad~C Coroamă} {and} \bibinfo{person}{Friedemann Mattern}.} \bibinfo{year}{2019}\natexlab{}.
\newblock \showarticletitle{Digital {Rebound} – {Why} {Digitalization} {Will} not {Redeem} us our {Environmental} {Sins}}. In \bibinfo{booktitle}{\emph{Proc. of the 6th {Int}. {Conf}. on {ICT} for {Sustainability} ({ICT4S})}}. \bibinfo{address}{Lappeenranta, Finland}.
\newblock
\urldef\tempurl%
\url{http://ceur-ws.org/Vol-2382/ICT4S2019_paper_31.pdf}
\showURL{%
\tempurl}


\bibitem[Coroamă and Pargman(2020)]%
        {coroama_skill_2020}
\bibfield{author}{\bibinfo{person}{Vlad~C. Coroamă} {and} \bibinfo{person}{Daniel Pargman}.} \bibinfo{year}{2020}\natexlab{}.
\newblock \showarticletitle{Skill rebound: {On} an unintended effect of digitalization}. In \bibinfo{booktitle}{\emph{Proceedings of the 7th {International} {Conference} on {ICT} for {Sustainability}}}. \bibinfo{publisher}{ACM}, \bibinfo{address}{Bristol United Kingdom}, \bibinfo{pages}{213--219}.
\newblock
\showISBNx{978-1-4503-7595-5}
\urldef\tempurl%
\url{https://doi.org/10.1145/3401335.3401362}
\showDOI{\tempurl}


\bibitem[Costanza et~al\mbox{.}(2014)]%
        {costanza2014doing}
\bibfield{author}{\bibinfo{person}{Enrico Costanza}, \bibinfo{person}{Joel~E Fischer}, \bibinfo{person}{James~A Colley}, \bibinfo{person}{Tom Rodden}, \bibinfo{person}{Sarvapali~D Ramchurn}, {and} \bibinfo{person}{Nicholas~R Jennings}.} \bibinfo{year}{2014}\natexlab{}.
\newblock \showarticletitle{Doing the laundry with agents: a field trial of a future smart energy system in the home}. In \bibinfo{booktitle}{\emph{Proceedings of the SIGCHI Conference on Human Factors in Computing Systems}}. \bibinfo{pages}{813--822}.
\newblock


\bibitem[Dankwa(2020)]%
        {dankwa2020investigating}
\bibfield{author}{\bibinfo{person}{Nana~Kesewaa Dankwa}.} \bibinfo{year}{2020}\natexlab{}.
\newblock \showarticletitle{Investigating Smart Home Needs for Elderly Women Who Live Alone. An Interview Study}. In \bibinfo{booktitle}{\emph{International Conference on Human-Computer Interaction}}. Springer, \bibinfo{pages}{32--38}.
\newblock


\bibitem[Darby(2018)]%
        {darby_smart_2018}
\bibfield{author}{\bibinfo{person}{Sarah~J. Darby}.} \bibinfo{year}{2018}\natexlab{}.
\newblock \showarticletitle{Smart technology in the home: time for more clarity}.
\newblock \bibinfo{journal}{\emph{Building Research \& Information}} \bibinfo{volume}{46}, \bibinfo{number}{1} (\bibinfo{date}{Jan.} \bibinfo{year}{2018}), \bibinfo{pages}{140--147}.
\newblock
\showISSN{0961-3218, 1466-4321}
\urldef\tempurl%
\url{https://doi.org/10.1080/09613218.2017.1301707}
\showDOI{\tempurl}


\bibitem[Dillahunt et~al\mbox{.}(2009)]%
        {dillahunt2009s}
\bibfield{author}{\bibinfo{person}{Tawanna Dillahunt}, \bibinfo{person}{Jennifer Mankoff}, \bibinfo{person}{Eric Paulos}, {and} \bibinfo{person}{Susan Fussell}.} \bibinfo{year}{2009}\natexlab{}.
\newblock \showarticletitle{It's not all about "Green": energy use in low-income communities}. In \bibinfo{booktitle}{\emph{Proceedings of the 11th international conference on Ubiquitous computing}}. \bibinfo{pages}{255--264}.
\newblock


\bibitem[DiSalvo et~al\mbox{.}(2010)]%
        {disalvo2010mapping}
\bibfield{author}{\bibinfo{person}{Carl DiSalvo}, \bibinfo{person}{Phoebe Sengers}, {and} \bibinfo{person}{Hr{\"o}nn Brynjarsd{\'o}ttir}.} \bibinfo{year}{2010}\natexlab{}.
\newblock \showarticletitle{Mapping the landscape of sustainable HCI}. In \bibinfo{booktitle}{\emph{Proceedings of the SIGCHI conference on human factors in computing systems}}. \bibinfo{pages}{1975--1984}.
\newblock


\bibitem[Duranton and Turner(2009)]%
        {duranton2009fundamental}
\bibfield{author}{\bibinfo{person}{Gilles Duranton} {and} \bibinfo{person}{Matthew~A Turner}.} \bibinfo{year}{2009}\natexlab{}.
\newblock \bibinfo{booktitle}{\emph{The fundamental law of road congestion: evidence from US cities}}.
\newblock \bibinfo{type}{{T}echnical {R}eport}. \bibinfo{institution}{National Bureau of Economic Research}.
\newblock


\bibitem[Easterbrook(2014)]%
        {easterbrook2014computational}
\bibfield{author}{\bibinfo{person}{Steve Easterbrook}.} \bibinfo{year}{2014}\natexlab{}.
\newblock \showarticletitle{From Computational Thinking to Systems Thinking: A conceptual toolkit for sustainability computing}. In \bibinfo{booktitle}{\emph{ICT for Sustainability 2014 (ICT4S-14)}}. Atlantis Press, \bibinfo{pages}{235--244}.
\newblock


\bibitem[{European Commission}(2015)]%
        {ECreport}
\bibfield{author}{\bibinfo{person}{{European Commission}}.} \bibinfo{year}{2015}\natexlab{}.
\newblock \bibinfo{title}{Towards an Integrated Strategic Energy Technology (SET) Plan: Accelerating the European Energy System Transformation}.
\newblock
\newblock


\bibitem[Fanning et~al\mbox{.}(2022)]%
        {fanning_social_2022}
\bibfield{author}{\bibinfo{person}{Andrew~L. Fanning}, \bibinfo{person}{Daniel~W. O’Neill}, \bibinfo{person}{Jason Hickel}, {and} \bibinfo{person}{Nicolas Roux}.} \bibinfo{year}{2022}\natexlab{}.
\newblock \showarticletitle{The social shortfall and ecological overshoot of nations {\textbar} {Nature} {Sustainability}}.
\newblock \bibinfo{journal}{\emph{Nature Sustainability}} \bibinfo{volume}{5}, \bibinfo{number}{1} (\bibinfo{date}{Jan.} \bibinfo{year}{2022}), \bibinfo{pages}{26--36}.
\newblock
\showISSN{2398-9629}
\urldef\tempurl%
\url{https://doi.org/10.1038/s41893-021-00799-z}
\showDOI{\tempurl}


\bibitem[Ford et~al\mbox{.}(2017)]%
        {ford_categories_2017}
\bibfield{author}{\bibinfo{person}{Rebecca Ford}, \bibinfo{person}{Marco Pritoni}, \bibinfo{person}{Angela Sanguinetti}, {and} \bibinfo{person}{Beth Karlin}.} \bibinfo{year}{2017}\natexlab{}.
\newblock \showarticletitle{Categories and functionality of smart home technology for energy management}.
\newblock \bibinfo{journal}{\emph{Building and Environment}}  \bibinfo{volume}{123} (\bibinfo{date}{Oct.} \bibinfo{year}{2017}), \bibinfo{pages}{543--554}.
\newblock
\showISSN{03601323}
\urldef\tempurl%
\url{https://doi.org/10.1016/j.buildenv.2017.07.020}
\showDOI{\tempurl}


\bibitem[Freitag et~al\mbox{.}(2021)]%
        {freitag2021climate}
\bibfield{author}{\bibinfo{person}{Charlotte Freitag}, \bibinfo{person}{Mike Berners-Lee}, \bibinfo{person}{Kelly Widdicks}, \bibinfo{person}{Bran Knowles}, \bibinfo{person}{Gordon Blair}, {and} \bibinfo{person}{Adrian Friday}.} \bibinfo{year}{2021}\natexlab{}.
\newblock \showarticletitle{The climate impact of ICT: A review of estimates, trends and regulations}.
\newblock \bibinfo{journal}{\emph{arXiv preprint arXiv:2102.02622}} (\bibinfo{year}{2021}).
\newblock


\bibitem[Gillingham et~al\mbox{.}(2013)]%
        {gillingham2013:rebound-overplayed}
\bibfield{author}{\bibinfo{person}{Kenneth Gillingham}, \bibinfo{person}{Matthew~J. Kotchen}, \bibinfo{person}{David~S. Rapson}, {and} \bibinfo{person}{Gernot Wagner}.} \bibinfo{year}{2013}\natexlab{}.
\newblock \showarticletitle{The rebound effect is overplayed}.
\newblock \bibinfo{journal}{\emph{Nature}} \bibinfo{volume}{493}, \bibinfo{number}{7433} (\bibinfo{date}{Jan.} \bibinfo{year}{2013}), \bibinfo{pages}{475--476}.
\newblock
\showISSN{0028-0836, 1476-4687}
\urldef\tempurl%
\url{https://doi.org/10.1038/493475a}
\showDOI{\tempurl}


\bibitem[Gram-Hanssen and Georg(2018)]%
        {gram-hanssen_energy_2018}
\bibfield{author}{\bibinfo{person}{Kirsten Gram-Hanssen} {and} \bibinfo{person}{Susse Georg}.} \bibinfo{year}{2018}\natexlab{}.
\newblock \showarticletitle{Energy performance gaps: promises, people, practices}.
\newblock \bibinfo{journal}{\emph{Building Research \& Information}} \bibinfo{volume}{46}, \bibinfo{number}{1} (\bibinfo{date}{Jan.} \bibinfo{year}{2018}), \bibinfo{pages}{1--9}.
\newblock
\showISSN{0961-3218, 1466-4321}
\urldef\tempurl%
\url{https://doi.org/10.1080/09613218.2017.1356127}
\showDOI{\tempurl}


\bibitem[Guerra~Santin(2013)]%
        {guerra2013occupant}
\bibfield{author}{\bibinfo{person}{Olivia Guerra~Santin}.} \bibinfo{year}{2013}\natexlab{}.
\newblock \showarticletitle{Occupant behaviour in energy efficient dwellings: evidence of a rebound effect}.
\newblock \bibinfo{journal}{\emph{Journal of Housing and the Built Environment}} \bibinfo{volume}{28}, \bibinfo{number}{2} (\bibinfo{year}{2013}), \bibinfo{pages}{311--327}.
\newblock


\bibitem[Haas and Biermayr(2000)]%
        {haas_rebound_2000}
\bibfield{author}{\bibinfo{person}{Reinhard Haas} {and} \bibinfo{person}{Peter Biermayr}.} \bibinfo{year}{2000}\natexlab{}.
\newblock \showarticletitle{The rebound effect for space heating {Empirical} evidence from {Austria}}.
\newblock \bibinfo{journal}{\emph{Energy Policy}} \bibinfo{volume}{28}, \bibinfo{number}{6-7} (\bibinfo{year}{2000}), \bibinfo{pages}{403--410}.
\newblock


\bibitem[Haberl et~al\mbox{.}(2020)]%
        {haberl_systematic_2020}
\bibfield{author}{\bibinfo{person}{Helmut Haberl}, \bibinfo{person}{Dominik Wiedenhofer}, \bibinfo{person}{Doris Virág}, \bibinfo{person}{Gerald Kalt}, \bibinfo{person}{Barbara Plank}, \bibinfo{person}{Paul Brockway}, \bibinfo{person}{Tomer Fishman}, \bibinfo{person}{Daniel Hausknost}, \bibinfo{person}{Fridolin Krausmann}, \bibinfo{person}{Bartholomäus Leon-Gruchalski}, \bibinfo{person}{Andreas Mayer}, \bibinfo{person}{Melanie Pichler}, \bibinfo{person}{Anke Schaffartzik}, \bibinfo{person}{Tânia Sousa}, \bibinfo{person}{Jan Streeck}, {and} \bibinfo{person}{Felix Creutzig}.} \bibinfo{year}{2020}\natexlab{}.
\newblock \showarticletitle{A systematic review of the evidence on decoupling of {GDP}, resource use and {GHG} emissions, part {II}: synthesizing the insights}.
\newblock \bibinfo{journal}{\emph{Environmental Research Letters}} \bibinfo{volume}{15}, \bibinfo{number}{6} (\bibinfo{date}{June} \bibinfo{year}{2020}), \bibinfo{pages}{065003}.
\newblock
\showISSN{1748-9326}
\urldef\tempurl%
\url{https://doi.org/10.1088/1748-9326/ab842a}
\showDOI{\tempurl}
\newblock
\shownote{Publisher: IOP Publishing}.


\bibitem[Hargreaves et~al\mbox{.}(2010)]%
        {hargreaves2010making}
\bibfield{author}{\bibinfo{person}{Tom Hargreaves}, \bibinfo{person}{Michael Nye}, {and} \bibinfo{person}{Jacquelin Burgess}.} \bibinfo{year}{2010}\natexlab{}.
\newblock \showarticletitle{Making energy visible: A qualitative field study of how householders interact with feedback from smart energy monitors}.
\newblock \bibinfo{journal}{\emph{Energy policy}} \bibinfo{volume}{38}, \bibinfo{number}{10} (\bibinfo{year}{2010}), \bibinfo{pages}{6111--6119}.
\newblock


\bibitem[Hargreaves et~al\mbox{.}(2018)]%
        {hargreaves_learning_2018}
\bibfield{author}{\bibinfo{person}{Tom Hargreaves}, \bibinfo{person}{Charlie Wilson}, {and} \bibinfo{person}{Richard Hauxwell-Baldwin}.} \bibinfo{year}{2018}\natexlab{}.
\newblock \showarticletitle{Learning to live in a smart home}.
\newblock \bibinfo{journal}{\emph{Building Research \& Information}} \bibinfo{volume}{46}, \bibinfo{number}{1} (\bibinfo{date}{Jan.} \bibinfo{year}{2018}), \bibinfo{pages}{127--139}.
\newblock
\showISSN{0961-3218, 1466-4321}
\urldef\tempurl%
\url{https://doi.org/10.1080/09613218.2017.1286882}
\showDOI{\tempurl}


\bibitem[Hens et~al\mbox{.}(2010)]%
        {hens2010energy}
\bibfield{author}{\bibinfo{person}{Hugo Hens}, \bibinfo{person}{Wout Parijs}, {and} \bibinfo{person}{Mieke Deurinck}.} \bibinfo{year}{2010}\natexlab{}.
\newblock \showarticletitle{Energy consumption for heating and rebound effects}.
\newblock \bibinfo{journal}{\emph{Energy and buildings}} \bibinfo{volume}{42}, \bibinfo{number}{1} (\bibinfo{year}{2010}), \bibinfo{pages}{105--110}.
\newblock


\bibitem[Hickel(2020)]%
        {hickel_less_2020}
\bibfield{author}{\bibinfo{person}{Jason Hickel}.} \bibinfo{year}{2020}\natexlab{}.
\newblock \bibinfo{booktitle}{\emph{Less is {More}: {How} {Degrowth} {Will} {Save} the {World}}}.
\newblock \bibinfo{publisher}{Penguin Random House}.
\newblock


\bibitem[Hickel and Kallis(2020)]%
        {hickel_is_2020}
\bibfield{author}{\bibinfo{person}{Jason Hickel} {and} \bibinfo{person}{Giorgos Kallis}.} \bibinfo{year}{2020}\natexlab{}.
\newblock \showarticletitle{Is {Green} {Growth} {Possible}?}
\newblock \bibinfo{journal}{\emph{New Political Economy}} \bibinfo{volume}{25}, \bibinfo{number}{4} (\bibinfo{date}{June} \bibinfo{year}{2020}), \bibinfo{pages}{469--486}.
\newblock
\showISSN{1356-3467}
\urldef\tempurl%
\url{https://doi.org/10.1080/13563467.2019.1598964}
\showDOI{\tempurl}
\newblock
\shownote{Publisher: Routledge \_eprint: https://doi.org/10.1080/13563467.2019.1598964}.


\bibitem[Hickel et~al\mbox{.}(2022)]%
        {hickel_national_2022}
\bibfield{author}{\bibinfo{person}{Jason Hickel}, \bibinfo{person}{Daniel~W. O’Neill}, \bibinfo{person}{Andrew~L. Fanning}, {and} \bibinfo{person}{Huzaifa Zoomkawala}.} \bibinfo{year}{2022}\natexlab{}.
\newblock \showarticletitle{National responsibility for ecological breakdown: a fair-shares assessment of resource use, 1970–2017}.
\newblock \bibinfo{journal}{\emph{The Lancet Planetary Health}} \bibinfo{volume}{6}, \bibinfo{number}{4} (\bibinfo{date}{April} \bibinfo{year}{2022}), \bibinfo{pages}{e342--e349}.
\newblock
\showISSN{2542-5196}
\urldef\tempurl%
\url{https://doi.org/10.1016/S2542-5196(22)00044-4}
\showDOI{\tempurl}
\newblock
\shownote{Publisher: Elsevier}.


\bibitem[Hilty et~al\mbox{.}(2011)]%
        {hilty2011sustainability}
\bibfield{author}{\bibinfo{person}{Lorenz Hilty}, \bibinfo{person}{Wolfgang Lohmann}, {and} \bibinfo{person}{Elaine~M Huang}.} \bibinfo{year}{2011}\natexlab{}.
\newblock \showarticletitle{Sustainability and ICT-an overview of the field}.
\newblock \bibinfo{journal}{\emph{Notizie di POLITEIA}} \bibinfo{volume}{27}, \bibinfo{number}{104} (\bibinfo{year}{2011}), \bibinfo{pages}{13--28}.
\newblock


\bibitem[Hilty and Aebischer(2015)]%
        {hilty2015ict}
\bibfield{author}{\bibinfo{person}{Lorenz~M Hilty} {and} \bibinfo{person}{Bernard Aebischer}.} \bibinfo{year}{2015}\natexlab{}.
\newblock \showarticletitle{ICT for sustainability: An emerging research field}.
\newblock \bibinfo{journal}{\emph{ICT innovations for Sustainability}} (\bibinfo{year}{2015}), \bibinfo{pages}{3--36}.
\newblock


\bibitem[Ilstedt and Wangel(2014)]%
        {ilstedt2014altering}
\bibfield{author}{\bibinfo{person}{Sara Ilstedt} {and} \bibinfo{person}{Josefin Wangel}.} \bibinfo{year}{2014}\natexlab{}.
\newblock \showarticletitle{Altering expectations: How design fictions and backcasting can leverage sustainable lifestyles}. In \bibinfo{booktitle}{\emph{DRS (Design Research Society) 2014: Design's Big Debates-Pushing the Boundaries of Design Research. Ume{\aa}, Sweden, June 16-19 2014}}.
\newblock


\bibitem[Jakobi et~al\mbox{.}(2017)]%
        {jakobi2017catch}
\bibfield{author}{\bibinfo{person}{Timo Jakobi}, \bibinfo{person}{Corinna Ogonowski}, \bibinfo{person}{Nico Castelli}, \bibinfo{person}{Gunnar Stevens}, {and} \bibinfo{person}{Volker Wulf}.} \bibinfo{year}{2017}\natexlab{}.
\newblock \showarticletitle{The catch(es) with smart home: Experiences of a living lab field study}. In \bibinfo{booktitle}{\emph{Proceedings of the 2017 CHI Conference on Human Factors in Computing Systems}}. \bibinfo{pages}{1620--1633}.
\newblock


\bibitem[Jensen et~al\mbox{.}(2018a)]%
        {jensen2018designing}
\bibfield{author}{\bibinfo{person}{Rikke~Hagensby Jensen}, \bibinfo{person}{Yolande Strengers}, \bibinfo{person}{Jesper Kjeldskov}, \bibinfo{person}{Larissa Nicholls}, {and} \bibinfo{person}{Mikael~B Skov}.} \bibinfo{year}{2018}\natexlab{a}.
\newblock \showarticletitle{Designing the desirable smart home: A study of household experiences and energy consumption impacts}. In \bibinfo{booktitle}{\emph{Proceedings of the 2018 CHI Conference on Human Factors in Computing Systems}}. \bibinfo{pages}{1--14}.
\newblock


\bibitem[Jensen et~al\mbox{.}(2018b)]%
        {jensen2018exploring}
\bibfield{author}{\bibinfo{person}{Rikke~Hagensby Jensen}, \bibinfo{person}{Yolande Strengers}, \bibinfo{person}{Dimitrios Raptis}, \bibinfo{person}{Larissa Nicholls}, \bibinfo{person}{Jesper Kjeldskov}, {and} \bibinfo{person}{Mikael~B Skov}.} \bibinfo{year}{2018}\natexlab{b}.
\newblock \showarticletitle{Exploring Hygge as a desirable design vision for the sustainable smart home}. In \bibinfo{booktitle}{\emph{Proceedings of the 2018 Designing Interactive Systems Conference}}. \bibinfo{pages}{355--360}.
\newblock


\bibitem[Jevons(1865)]%
        {jevons1865coal}
\bibfield{author}{\bibinfo{person}{William~Stanley Jevons}.} \bibinfo{year}{1865}\natexlab{}.
\newblock \bibinfo{booktitle}{\emph{The Coal Question; An Inquiry Concerning the Progress of the Nation, and the Probable Exhaustion of Our Coal-mines}}.
\newblock \bibinfo{publisher}{Macmillan and Co.}
\newblock


\bibitem[Kallman and Frickel(2019)]%
        {kallman2019power}
\bibfield{author}{\bibinfo{person}{Meghan~Elizabeth Kallman} {and} \bibinfo{person}{Scott Frickel}.} \bibinfo{year}{2019}\natexlab{}.
\newblock \showarticletitle{Power to the people: industrial transition movements and energy populism}.
\newblock \bibinfo{journal}{\emph{Environmental Sociology}} \bibinfo{volume}{5}, \bibinfo{number}{3} (\bibinfo{year}{2019}), \bibinfo{pages}{255--268}.
\newblock


\bibitem[Kaufman and Silberman(2011)]%
        {kaufman2011rebound}
\bibfield{author}{\bibinfo{person}{Samuel~J Kaufman} {and} \bibinfo{person}{M~Six Silberman}.} \bibinfo{year}{2011}\natexlab{}.
\newblock \showarticletitle{Rebound effects in sustainable HCI}. In \bibinfo{booktitle}{\emph{Sustainable Interaction Design in Professional Domains, workshop at CHI 2011}}. Citeseer.
\newblock


\bibitem[Khazzoom(1980)]%
        {khazzoom1980:rebound}
\bibfield{author}{\bibinfo{person}{J.~Daniel Khazzoom}.} \bibinfo{year}{1980}\natexlab{}.
\newblock \showarticletitle{Economic {Implications} of {Mandated} {Efficiency} in {Standards} for {Household} {Appliances}}.
\newblock \bibinfo{journal}{\emph{The Energy Journal}} \bibinfo{volume}{1}, \bibinfo{number}{4} (\bibinfo{year}{1980}), \bibinfo{pages}{21--40}.
\newblock
\showISSN{0195-6574}
\urldef\tempurl%
\url{https://www.jstor.org/stable/41321476}
\showURL{%
\tempurl}
\newblock
\shownote{Publisher: International Association for Energy Economics}.


\bibitem[Kisselburgh et~al\mbox{.}(2020)]%
        {kisselburgh2020hci}
\bibfield{author}{\bibinfo{person}{Lorraine Kisselburgh}, \bibinfo{person}{Michel Beaudouin-Lafon}, \bibinfo{person}{Lorrie Cranor}, \bibinfo{person}{Jonathan Lazar}, {and} \bibinfo{person}{Vicki~L Hanson}.} \bibinfo{year}{2020}\natexlab{}.
\newblock \showarticletitle{HCI ethics, privacy, accessibility, and the environment: A town hall forum on global policy issues}. In \bibinfo{booktitle}{\emph{Extended Abstracts of the 2020 CHI Conference on Human Factors in Computing Systems}}. \bibinfo{pages}{1--6}.
\newblock


\bibitem[Kitchenham et~al\mbox{.}(2009)]%
        {kitchenham_systematic_2009}
\bibfield{author}{\bibinfo{person}{Barbara Kitchenham}, \bibinfo{person}{O.~Pearl Brereton}, \bibinfo{person}{David Budgen}, \bibinfo{person}{Mark Turner}, \bibinfo{person}{John Bailey}, {and} \bibinfo{person}{Stephen Linkman}.} \bibinfo{year}{2009}\natexlab{}.
\newblock \showarticletitle{Systematic literature reviews in software engineering–a systematic literature review}.
\newblock \bibinfo{journal}{\emph{Information and software technology}} \bibinfo{volume}{51}, \bibinfo{number}{1} (\bibinfo{year}{2009}).
\newblock
\urldef\tempurl%
\url{http://www.sciencedirect.com/science/article/pii/S0950584908001390}
\showURL{%
\tempurl}


\bibitem[Knowles(2014)]%
        {knowles_cyber-sustainabilitytowards_2013}
\bibfield{author}{\bibinfo{person}{Bran Knowles}.} \bibinfo{year}{2014}\natexlab{}.
\newblock \emph{\bibinfo{title}{Cyber-sustainability: towards a sustainable digital future}}.
\newblock {PhD} thesis. \bibinfo{school}{Lancaster University}, \bibinfo{address}{Lancaster}.
\newblock
\urldef\tempurl%
\url{http://eprints.lancs.ac.uk/68468/}
\showURL{%
\tempurl}


\bibitem[Koomey et~al\mbox{.}(2011)]%
        {koomey2011:law}
\bibfield{author}{\bibinfo{person}{Jonathan Koomey}, \bibinfo{person}{Stephen Berard}, \bibinfo{person}{Marla Sanchez}, {and} \bibinfo{person}{Henry Wong}.} \bibinfo{year}{2011}\natexlab{}.
\newblock \showarticletitle{Implications of Historical Trends in the Electrical Efficiency of Computing}.
\newblock \bibinfo{journal}{\emph{IEEE Annals of the History of Computing}} \bibinfo{volume}{33}, \bibinfo{number}{3} (\bibinfo{year}{2011}), \bibinfo{pages}{46--54}.
\newblock
\urldef\tempurl%
\url{https://doi.org/10.1109/MAHC.2010.28}
\showDOI{\tempurl}


\bibitem[Kumar and Mani(2017)]%
        {kumar_energy-neutrality_2017}
\bibfield{author}{\bibinfo{person}{Tarun Kumar} {and} \bibinfo{person}{Monto Mani}.} \bibinfo{year}{2017}\natexlab{}.
\newblock \showarticletitle{An {Energy}-{Neutrality} based {Evaluation} into the {Effectiveness} of {Occupancy} {Sensors} in {Buildings}: {An} integrated {Life}-cycle {Study}}. In \bibinfo{booktitle}{\emph{Proceedings of 33rd {PLEA} {International} {Conference}: {Design} to {Thrive}, {PLEA} 2017}}. \bibinfo{publisher}{NCEUB 2017 - Network for Comfort and Energy Use in Buildings}, \bibinfo{address}{Edinburgh}, \bibinfo{pages}{2579--2586}.
\newblock
\showISBNx{978-0-9928957-5-4}


\bibitem[Lange et~al\mbox{.}(2019)]%
        {lange2019economy}
\bibfield{author}{\bibinfo{person}{Steffen Lange}, \bibinfo{person}{Maximilian Banning}, \bibinfo{person}{Anne Berner}, \bibinfo{person}{Florian Kern}, \bibinfo{person}{Christian Lutz}, \bibinfo{person}{Jan Peuckert}, \bibinfo{person}{Tilman Santarius}, \bibinfo{person}{Alexander Silbersdorff}, {and} \bibinfo{person}{ReCap Arbeitsbericht}.} \bibinfo{year}{2019}\natexlab{}.
\newblock \showarticletitle{Economy-Wide Rebound Effects: State of the art, a new taxonomy, policy and research gaps}.
\newblock \bibinfo{journal}{\emph{ReCap Makro-Rebounds begrenzen}} (\bibinfo{year}{2019}).
\newblock


\bibitem[Lennon(2021)]%
        {lennon2021:energy-justice}
\bibfield{author}{\bibinfo{person}{Myles Lennon}.} \bibinfo{year}{2021}\natexlab{}.
\newblock \showarticletitle{Energy transitions in a time of intersecting precarities: {From} reductive environmentalism to antiracist praxis}.
\newblock \bibinfo{journal}{\emph{Energy Research \& Social Science}}  \bibinfo{volume}{73} (\bibinfo{date}{March} \bibinfo{year}{2021}).
\newblock
\showISSN{2214-6296}
\urldef\tempurl%
\url{https://doi.org/10.1016/j.erss.2021.101930}
\showDOI{\tempurl}


\bibitem[Lewis(1973)]%
        {lewis1973anthropology}
\bibfield{author}{\bibinfo{person}{Diane Lewis}.} \bibinfo{year}{1973}\natexlab{}.
\newblock \showarticletitle{Anthropology and colonialism}.
\newblock \bibinfo{journal}{\emph{Current anthropology}} \bibinfo{volume}{14}, \bibinfo{number}{5} (\bibinfo{year}{1973}), \bibinfo{pages}{581--602}.
\newblock


\bibitem[{Li Jiang} et~al\mbox{.}(2004)]%
        {li_jiang_smart_2004}
\bibfield{author}{\bibinfo{person}{{Li Jiang}}, \bibinfo{person}{{Da-You Liu}}, {and} \bibinfo{person}{{Bo Yang}}.} \bibinfo{year}{2004}\natexlab{}.
\newblock \showarticletitle{Smart home research}. In \bibinfo{booktitle}{\emph{Proceedings of 2004 {International} {Conference} on {Machine} {Learning} and {Cybernetics} ({IEEE} {Cat}. {No}.{04EX826})}}, Vol.~\bibinfo{volume}{2}. \bibinfo{publisher}{IEEE}, \bibinfo{address}{Shanghai, China}, \bibinfo{pages}{659--663}.
\newblock
\showISBNx{978-0-7803-8403-3}
\urldef\tempurl%
\url{https://doi.org/10.1109/ICMLC.2004.1382266}
\showDOI{\tempurl}


\bibitem[Lutolf(1992)]%
        {lutolf1992smart}
\bibfield{author}{\bibinfo{person}{Remo Lutolf}.} \bibinfo{year}{1992}\natexlab{}.
\newblock \showarticletitle{Smart home concept and the integration of energy meters into a home based system}. In \bibinfo{booktitle}{\emph{Seventh international conference on metering apparatus and tariffs for electricity supply 1992}}. IET, \bibinfo{pages}{277--278}.
\newblock


\bibitem[Mann et~al\mbox{.}(2018)]%
        {mann_shifting_2018}
\bibfield{author}{\bibinfo{person}{Samuel Mann}, \bibinfo{person}{Oliver Bates}, {and} \bibinfo{person}{Raymond Maher}.} \bibinfo{year}{2018}\natexlab{}.
\newblock \showarticletitle{Shifting the maturity needle of {ICT} for {Sustainability}}.
\newblock \bibinfo{journal}{\emph{ICT4S2018. 5th International Conference on Information and Communication Technology for Sustainability}} (\bibinfo{year}{2018}), \bibinfo{pages}{209--226}.
\newblock
\urldef\tempurl%
\url{https://doi.org/10.29007/d6g3}
\showDOI{\tempurl}


\bibitem[Milchram et~al\mbox{.}(2018)]%
        {milchram2018energy}
\bibfield{author}{\bibinfo{person}{Christine Milchram}, \bibinfo{person}{Rafaela Hillerbrand}, \bibinfo{person}{Geerten van~de Kaa}, \bibinfo{person}{Neelke Doorn}, {and} \bibinfo{person}{Rolf K{\"u}nneke}.} \bibinfo{year}{2018}\natexlab{}.
\newblock \showarticletitle{Energy justice and smart grid systems: evidence from the Netherlands and the United Kingdom}.
\newblock \bibinfo{journal}{\emph{Applied Energy}}  \bibinfo{volume}{229} (\bibinfo{year}{2018}), \bibinfo{pages}{1244--1259}.
\newblock


\bibitem[Nardi(2019)]%
        {nardi_design_2019}
\bibfield{author}{\bibinfo{person}{Bonnie Nardi}.} \bibinfo{year}{2019}\natexlab{}.
\newblock \showarticletitle{Design in the {Age} of {Climate} {Change}}.
\newblock \bibinfo{journal}{\emph{She Ji: The Journal of Design, Economics, and Innovation}} \bibinfo{volume}{5}, \bibinfo{number}{1} (\bibinfo{date}{March} \bibinfo{year}{2019}), \bibinfo{pages}{5--14}.
\newblock
\showISSN{2405-8726}
\urldef\tempurl%
\url{https://doi.org/10.1016/j.sheji.2019.01.001}
\showDOI{\tempurl}


\bibitem[Nathan et~al\mbox{.}(2008)]%
        {nathan2008envisioning}
\bibfield{author}{\bibinfo{person}{Lisa~P Nathan}, \bibinfo{person}{Batya Friedman}, \bibinfo{person}{Predrag Klasnja}, \bibinfo{person}{Shaun~K Kane}, {and} \bibinfo{person}{Jessica~K Miller}.} \bibinfo{year}{2008}\natexlab{}.
\newblock \showarticletitle{Envisioning systemic effects on persons and society throughout interactive system design}. In \bibinfo{booktitle}{\emph{Proceedings of the 7th ACM conference on Designing interactive systems}}. \bibinfo{pages}{1--10}.
\newblock


\bibitem[Naylor et~al\mbox{.}(2018)]%
        {naylor_review_2018}
\bibfield{author}{\bibinfo{person}{Sophie Naylor}, \bibinfo{person}{Mark Gillott}, {and} \bibinfo{person}{Tom Lau}.} \bibinfo{year}{2018}\natexlab{}.
\newblock \showarticletitle{A review of occupant-centric building control strategies to reduce building energy use}.
\newblock \bibinfo{journal}{\emph{Renewable and Sustainable Energy Reviews}}  \bibinfo{volume}{96} (\bibinfo{date}{Nov.} \bibinfo{year}{2018}), \bibinfo{pages}{1--10}.
\newblock
\showISSN{13640321}
\urldef\tempurl%
\url{https://doi.org/10.1016/j.rser.2018.07.019}
\showDOI{\tempurl}


\bibitem[Nässén and Holmberg(2009)]%
        {nassen_quantifying_2009}
\bibfield{author}{\bibinfo{person}{Jonas Nässén} {and} \bibinfo{person}{John Holmberg}.} \bibinfo{year}{2009}\natexlab{}.
\newblock \showarticletitle{Quantifying the rebound effects of energy efficiency improvements and energy conserving behaviour in {Sweden}}.
\newblock \bibinfo{journal}{\emph{Energy Efficiency}} \bibinfo{volume}{2}, \bibinfo{number}{3} (\bibinfo{date}{Aug.} \bibinfo{year}{2009}), \bibinfo{pages}{221--231}.
\newblock
\showISSN{1570-646X, 1570-6478}
\urldef\tempurl%
\url{https://doi.org/10.1007/s12053-009-9046-x}
\showDOI{\tempurl}


\bibitem[Pargman and Raghavan(2014a)]%
        {pargman2014rethinking}
\bibfield{author}{\bibinfo{person}{Daniel Pargman} {and} \bibinfo{person}{Barath Raghavan}.} \bibinfo{year}{2014}\natexlab{a}.
\newblock \showarticletitle{Rethinking sustainability in computing: From buzzword to non-negotiable limits}. In \bibinfo{booktitle}{\emph{Proceedings of the 8th Nordic Conference on Human-Computer Interaction: Fun, Fast, Foundational}}. \bibinfo{pages}{638--647}.
\newblock


\bibitem[Pargman and Raghavan(2014b)]%
        {pargman2014:rebound}
\bibfield{author}{\bibinfo{person}{Daniel Pargman} {and} \bibinfo{person}{Barath Raghavan}.} \bibinfo{year}{2014}\natexlab{b}.
\newblock \showarticletitle{Rethinking Sustainability in Computing: From Buzzword to Non-Negotiable Limits} \emph{(\bibinfo{series}{NordiCHI '14})}. \bibinfo{publisher}{Association for Computing Machinery}, \bibinfo{address}{New York, NY, USA}, \bibinfo{pages}{638--–647}.
\newblock
\showISBNx{9781450325424}
\urldef\tempurl%
\url{https://doi.org/10.1145/2639189.2639228}
\showDOI{\tempurl}


\bibitem[Parrique et~al\mbox{.}(2019)]%
        {t_parrique_decoupling_2019}
\bibfield{author}{\bibinfo{person}{T. Parrique}, \bibinfo{person}{J. Barth}, \bibinfo{person}{F. Briens}, \bibinfo{person}{C. Kerschner}, \bibinfo{person}{A. Kraus-Polk}, \bibinfo{person}{A. Kuokkanen}, {and} \bibinfo{person}{J.H. Spangenberg}.} \bibinfo{year}{2019}\natexlab{}.
\newblock \bibinfo{booktitle}{\emph{Decoupling debunked – {Evidence} and arguments against green growth as a sole strategy for sustainability}}.
\newblock \bibinfo{publisher}{European Environmental Bureau}.
\newblock
\urldef\tempurl%
\url{https://eeb.org/library/decoupling-debunked/}
\showURL{%
\tempurl}


\bibitem[Penzenstadler et~al\mbox{.}(2018)]%
        {penzenstadler2018software}
\bibfield{author}{\bibinfo{person}{Birgit Penzenstadler}, \bibinfo{person}{Leticia Duboc}, \bibinfo{person}{Colin~C Venters}, \bibinfo{person}{Stefanie Betz}, \bibinfo{person}{Norbert Seyff}, \bibinfo{person}{Krzsztof Wnuk}, \bibinfo{person}{Ruzanna Chitchyan}, \bibinfo{person}{Steve~M Easterbrook}, {and} \bibinfo{person}{Christoph Becker}.} \bibinfo{year}{2018}\natexlab{}.
\newblock \showarticletitle{Software engineering for sustainability: Find the leverage points!}
\newblock \bibinfo{journal}{\emph{IEEE Software}} \bibinfo{volume}{35}, \bibinfo{number}{4} (\bibinfo{year}{2018}), \bibinfo{pages}{22--33}.
\newblock


\bibitem[Penzenstadler et~al\mbox{.}(2014)]%
        {penzenstadler2014safety}
\bibfield{author}{\bibinfo{person}{Birgit Penzenstadler}, \bibinfo{person}{Ankita Raturi}, \bibinfo{person}{Debra Richardson}, {and} \bibinfo{person}{Bill Tomlinson}.} \bibinfo{year}{2014}\natexlab{}.
\newblock \showarticletitle{Safety, security, now sustainability: The nonfunctional requirement for the 21st century}.
\newblock \bibinfo{journal}{\emph{IEEE software}} \bibinfo{volume}{31}, \bibinfo{number}{3} (\bibinfo{year}{2014}), \bibinfo{pages}{40--47}.
\newblock


\bibitem[Petersen et~al\mbox{.}(2008)]%
        {petersen2008systematic}
\bibfield{author}{\bibinfo{person}{Kai Petersen}, \bibinfo{person}{Robert Feldt}, \bibinfo{person}{Shahid Mujtaba}, {and} \bibinfo{person}{Michael Mattsson}.} \bibinfo{year}{2008}\natexlab{}.
\newblock \showarticletitle{Systematic mapping studies in software engineering}. In \bibinfo{booktitle}{\emph{12th International Conference on Evaluation and Assessment in Software Engineering (EASE) 12}}. \bibinfo{pages}{1--10}.
\newblock


\bibitem[Petersen et~al\mbox{.}(2015)]%
        {petersen_guidelines_2015}
\bibfield{author}{\bibinfo{person}{Kai Petersen}, \bibinfo{person}{Sairam Vakkalanka}, {and} \bibinfo{person}{Ludwik Kuzniarz}.} \bibinfo{year}{2015}\natexlab{}.
\newblock \showarticletitle{Guidelines for conducting systematic mapping studies in software engineering: {An} update}.
\newblock \bibinfo{journal}{\emph{Information and Software Technology}}  \bibinfo{volume}{64} (\bibinfo{date}{Aug.} \bibinfo{year}{2015}), \bibinfo{pages}{1--18}.
\newblock
\showISSN{0950-5849}
\urldef\tempurl%
\url{https://doi.org/10.1016/j.infsof.2015.03.007}
\showDOI{\tempurl}


\bibitem[Pierce(2012)]%
        {pierce2012undesigning}
\bibfield{author}{\bibinfo{person}{James Pierce}.} \bibinfo{year}{2012}\natexlab{}.
\newblock \showarticletitle{Undesigning technology: considering the negation of design by design}. In \bibinfo{booktitle}{\emph{Proceedings of the SIGCHI Conference on Human Factors in Computing Systems}}. \bibinfo{pages}{957--966}.
\newblock


\bibitem[Pierce and Paulos(2012)]%
        {pierce2012beyond}
\bibfield{author}{\bibinfo{person}{James Pierce} {and} \bibinfo{person}{Eric Paulos}.} \bibinfo{year}{2012}\natexlab{}.
\newblock \showarticletitle{Beyond energy monitors: interaction, energy, and emerging energy systems}. In \bibinfo{booktitle}{\emph{Proceedings of the SIGCHI Conference on Human Factors in Computing Systems}}. \bibinfo{pages}{665--674}.
\newblock


\bibitem[Pierce et~al\mbox{.}(2010)]%
        {Schiano2010}
\bibfield{author}{\bibinfo{person}{James Pierce}, \bibinfo{person}{Diane~J. Schiano}, {and} \bibinfo{person}{Eric Paulos}.} \bibinfo{year}{2010}\natexlab{}.
\newblock \showarticletitle{Home, Habits, and Energy: Examining Domestic Interactions and Energy Consumption}. In \bibinfo{booktitle}{\emph{Proceedings of the SIGCHI Conference on Human Factors in Computing Systems}} (Atlanta, Georgia, USA) \emph{(\bibinfo{series}{CHI '10})}. \bibinfo{publisher}{Association for Computing Machinery}, \bibinfo{address}{New York, NY, USA}, \bibinfo{pages}{1985–1994}.
\newblock
\showISBNx{9781605589299}
\urldef\tempurl%
\url{https://doi.org/10.1145/1753326.1753627}
\showDOI{\tempurl}


\bibitem[Podgornik et~al\mbox{.}(2013)]%
        {podgornik2013impact}
\bibfield{author}{\bibinfo{person}{Ales Podgornik}, \bibinfo{person}{Boris Sucic}, \bibinfo{person}{Peter Bevk}, {and} \bibinfo{person}{Damir Stanicic}.} \bibinfo{year}{2013}\natexlab{}.
\newblock \showarticletitle{The impact of smart metering on energy efficiency in low-income housing in Mediterranean}. In \bibinfo{booktitle}{\emph{Climate-Smart Technologies: Integrating Renewable Energy and Energy Efficiency in Mitigation and Adaptation Responses}}. Springer, \bibinfo{pages}{597--614}.
\newblock


\bibitem[Pohl et~al\mbox{.}(2019)]%
        {pohl2019:indirect-lca}
\bibfield{author}{\bibinfo{person}{Johanna Pohl}, \bibinfo{person}{Lorenz~M. Hilty}, {and} \bibinfo{person}{Matthias Finkbeiner}.} \bibinfo{year}{2019}\natexlab{}.
\newblock \showarticletitle{How {LCA} contributes to the environmental assessment of higher order effects of {ICT} application: {A} review of different approaches}.
\newblock \bibinfo{journal}{\emph{Journal of Cleaner Production}}  \bibinfo{volume}{219} (\bibinfo{date}{May} \bibinfo{year}{2019}), \bibinfo{pages}{698--712}.
\newblock
\showISSN{0959-6526}
\urldef\tempurl%
\url{https://doi.org/10.1016/j.jclepro.2019.02.018}
\showDOI{\tempurl}


\bibitem[Preist et~al\mbox{.}(2016)]%
        {preist2016understanding}
\bibfield{author}{\bibinfo{person}{Chris Preist}, \bibinfo{person}{Daniel Schien}, {and} \bibinfo{person}{Eli Blevis}.} \bibinfo{year}{2016}\natexlab{}.
\newblock \showarticletitle{Understanding and mitigating the effects of device and cloud service design decisions on the environmental footprint of digital infrastructure}. In \bibinfo{booktitle}{\emph{Proceedings of the 2016 CHI Conference on Human Factors in Computing Systems}}. \bibinfo{pages}{1324--1337}.
\newblock


\bibitem[Pérez-Lombard et~al\mbox{.}(2008)]%
        {perez-lombard_review_2008}
\bibfield{author}{\bibinfo{person}{Luis Pérez-Lombard}, \bibinfo{person}{José Ortiz}, {and} \bibinfo{person}{Christine Pout}.} \bibinfo{year}{2008}\natexlab{}.
\newblock \showarticletitle{A review on buildings energy consumption information}.
\newblock \bibinfo{journal}{\emph{Energy and Buildings}} \bibinfo{volume}{40}, \bibinfo{number}{3} (\bibinfo{date}{Jan.} \bibinfo{year}{2008}), \bibinfo{pages}{394--398}.
\newblock
\showISSN{03787788}
\urldef\tempurl%
\url{https://doi.org/10.1016/j.enbuild.2007.03.007}
\showDOI{\tempurl}


\bibitem[Reinisch et~al\mbox{.}(2011)]%
        {reinisch2011thinkhome}
\bibfield{author}{\bibinfo{person}{Christian Reinisch}, \bibinfo{person}{MarioJ Kofler}, \bibinfo{person}{F{\'e}lix Iglesias}, {and} \bibinfo{person}{Wolfgang Kastner}.} \bibinfo{year}{2011}\natexlab{}.
\newblock \showarticletitle{Thinkhome energy efficiency in future smart homes}.
\newblock \bibinfo{journal}{\emph{EURASIP Journal on Embedded Systems}}  \bibinfo{volume}{2011} (\bibinfo{year}{2011}), \bibinfo{pages}{1--18}.
\newblock


\bibitem[Remy et~al\mbox{.}(2018)]%
        {remy2018evaluation}
\bibfield{author}{\bibinfo{person}{Christian Remy}, \bibinfo{person}{Oliver Bates}, \bibinfo{person}{Alan Dix}, \bibinfo{person}{Vanessa Thomas}, \bibinfo{person}{Mike Hazas}, \bibinfo{person}{Adrian Friday}, {and} \bibinfo{person}{Elaine~M Huang}.} \bibinfo{year}{2018}\natexlab{}.
\newblock \showarticletitle{Evaluation beyond usability: Validating sustainable HCI research}. In \bibinfo{booktitle}{\emph{Proceedings of the 2018 CHI Conference on Human Factors in Computing Systems}}. \bibinfo{pages}{1--14}.
\newblock


\bibitem[Rodden et~al\mbox{.}(2013)]%
        {rodden2013home}
\bibfield{author}{\bibinfo{person}{Tom~A Rodden}, \bibinfo{person}{Joel~E Fischer}, \bibinfo{person}{Nadia Pantidi}, \bibinfo{person}{Khaled Bachour}, {and} \bibinfo{person}{Stuart Moran}.} \bibinfo{year}{2013}\natexlab{}.
\newblock \showarticletitle{At home with agents: exploring attitudes towards future smart energy infrastructures}. In \bibinfo{booktitle}{\emph{Proceedings of the SIGCHI Conference on Human Factors in Computing Systems}}. \bibinfo{pages}{1173--1182}.
\newblock


\bibitem[Rogers and Marshall(2017)]%
        {rogers2017:wild}
\bibfield{author}{\bibinfo{person}{Yvonne Rogers} {and} \bibinfo{person}{Paul Marshall}.} \bibinfo{year}{2017}\natexlab{}.
\newblock \showarticletitle{Research in the Wild}.
\newblock \bibinfo{journal}{\emph{Synthesis Lectures on Human-Centered Informatics}} \bibinfo{volume}{10}, \bibinfo{number}{3} (\bibinfo{year}{2017}), \bibinfo{pages}{i--97}.
\newblock
\urldef\tempurl%
\url{https://doi.org/10.2200/S00764ED1V01Y201703HCI037}
\showDOI{\tempurl}


\bibitem[Safarzadeh et~al\mbox{.}(2020)]%
        {safarzadeh2020review}
\bibfield{author}{\bibinfo{person}{Soroush Safarzadeh}, \bibinfo{person}{Morteza Rasti-Barzoki}, {and} \bibinfo{person}{Seyed~Reza Hejazi}.} \bibinfo{year}{2020}\natexlab{}.
\newblock \showarticletitle{A review of optimal energy policy instruments on industrial energy efficiency programs, rebound effects, and government policies}.
\newblock \bibinfo{journal}{\emph{Energy Policy}}  \bibinfo{volume}{139} (\bibinfo{year}{2020}), \bibinfo{pages}{111342}.
\newblock


\bibitem[Santarius(2017)]%
        {santarius2017}
\bibfield{author}{\bibinfo{person}{Tilman Santarius}.} \bibinfo{year}{2017}\natexlab{}.
\newblock \bibinfo{title}{Digitalization, Efficiency and the Rebound Effect}.
\newblock
\newblock
\urldef\tempurl%
\url{https://www.degrowth.info/blog/digitalization-efficiency-and-the-rebound-effect}
\showURL{%
\tempurl}


\bibitem[Saunders(1992)]%
        {saunders1992:khazzoom-brookes}
\bibfield{author}{\bibinfo{person}{Harry~D. Saunders}.} \bibinfo{year}{1992}\natexlab{}.
\newblock \showarticletitle{The {Khazzoom}-{Brookes} {Postulate} and {Neoclassical} {Growth}}.
\newblock \bibinfo{journal}{\emph{The Energy Journal}} \bibinfo{volume}{13}, \bibinfo{number}{4} (\bibinfo{year}{1992}), \bibinfo{pages}{131--148}.
\newblock
\showISSN{0195-6574}
\urldef\tempurl%
\url{https://www.jstor.org/stable/41322471}
\showURL{%
\tempurl}
\newblock
\shownote{Publisher: International Association for Energy Economics}.


\bibitem[Saunders(2000)]%
        {saunders2000:rebound}
\bibfield{author}{\bibinfo{person}{Harry~D. Saunders}.} \bibinfo{year}{2000}\natexlab{}.
\newblock \showarticletitle{A view from the macro side: rebound, backfire, and Khazzoom–Brookes}.
\newblock \bibinfo{journal}{\emph{Energy Policy}} \bibinfo{volume}{28}, \bibinfo{number}{6} (\bibinfo{year}{2000}), \bibinfo{pages}{439--449}.
\newblock
\showISSN{0301-4215}
\urldef\tempurl%
\url{https://doi.org/10.1016/S0301-4215(00)00024-0}
\showDOI{\tempurl}


\bibitem[Saunders(2013)]%
        {saunders2013historical}
\bibfield{author}{\bibinfo{person}{Harry~D Saunders}.} \bibinfo{year}{2013}\natexlab{}.
\newblock \showarticletitle{Historical evidence for energy efficiency rebound in 30 US sectors and a toolkit for rebound analysts}.
\newblock \bibinfo{journal}{\emph{Technological Forecasting and Social Change}} \bibinfo{volume}{80}, \bibinfo{number}{7} (\bibinfo{year}{2013}), \bibinfo{pages}{1317--1330}.
\newblock


\bibitem[Saunders and Tsao(2012)]%
        {saunders_rebound_2012}
\bibfield{author}{\bibinfo{person}{Harry~D. Saunders} {and} \bibinfo{person}{Jeffrey~Y. Tsao}.} \bibinfo{year}{2012}\natexlab{}.
\newblock \showarticletitle{Rebound effects for lighting}.
\newblock \bibinfo{journal}{\emph{Energy Policy}}  \bibinfo{volume}{49} (\bibinfo{date}{Oct.} \bibinfo{year}{2012}), \bibinfo{pages}{477--478}.
\newblock
\showISSN{0301-4215}
\urldef\tempurl%
\url{https://doi.org/10.1016/j.enpol.2012.06.050}
\showDOI{\tempurl}


\bibitem[Seebauer(2018)]%
        {seebauer2018psychology}
\bibfield{author}{\bibinfo{person}{Sebastian Seebauer}.} \bibinfo{year}{2018}\natexlab{}.
\newblock \showarticletitle{The psychology of rebound effects: Explaining energy efficiency rebound behaviours with electric vehicles and building insulation in Austria}.
\newblock \bibinfo{journal}{\emph{Energy research \& social science}}  \bibinfo{volume}{46} (\bibinfo{year}{2018}), \bibinfo{pages}{311--320}.
\newblock


\bibitem[Soden et~al\mbox{.}(2021)]%
        {soden2021we}
\bibfield{author}{\bibinfo{person}{Robert Soden}, \bibinfo{person}{Pradnaya Pathak}, {and} \bibinfo{person}{Olivia Doggett}.} \bibinfo{year}{2021}\natexlab{}.
\newblock \showarticletitle{What we speculate about when we speculate about sustainable HCI}. In \bibinfo{booktitle}{\emph{ACM SIGCAS Conference on Computing and Sustainable Societies}}. \bibinfo{pages}{188--198}.
\newblock


\bibitem[Sorrell(2009)]%
        {sorrell_jevons_2009}
\bibfield{author}{\bibinfo{person}{Steve Sorrell}.} \bibinfo{year}{2009}\natexlab{}.
\newblock \showarticletitle{Jevons’ {Paradox} revisited: {The} evidence for backfire from improved energy efficiency}.
\newblock \bibinfo{journal}{\emph{Energy Policy}} \bibinfo{volume}{37}, \bibinfo{number}{4} (\bibinfo{date}{April} \bibinfo{year}{2009}), \bibinfo{pages}{1456--1469}.
\newblock
\showISSN{03014215}
\urldef\tempurl%
\url{https://doi.org/10.1016/j.enpol.2008.12.003}
\showDOI{\tempurl}


\bibitem[Sovacool et~al\mbox{.}(2023)]%
        {sovacool2023:pluralizing-justice}
\bibfield{author}{\bibinfo{person}{Benjamin~K. Sovacool}, \bibinfo{person}{Shannon~Elizabeth Bell}, \bibinfo{person}{Cara Daggett}, \bibinfo{person}{Christine Labuski}, \bibinfo{person}{Myles Lennon}, \bibinfo{person}{Lindsay Naylor}, \bibinfo{person}{Julie Klinger}, \bibinfo{person}{Kelsey Leonard}, {and} \bibinfo{person}{Jeremy Firestone}.} \bibinfo{year}{2023}\natexlab{}.
\newblock \showarticletitle{Pluralizing energy justice: {Incorporating} feminist, anti-racist, {Indigenous}, and postcolonial perspectives}.
\newblock \bibinfo{journal}{\emph{Energy Research \& Social Science}}  \bibinfo{volume}{97} (\bibinfo{date}{March} \bibinfo{year}{2023}).
\newblock
\showISSN{2214-6296}
\urldef\tempurl%
\url{https://doi.org/10.1016/j.erss.2023.102996}
\showDOI{\tempurl}


\bibitem[Ssozi-Mugarura et~al\mbox{.}(2016)]%
        {ssozi2016enough}
\bibfield{author}{\bibinfo{person}{Fiona Ssozi-Mugarura}, \bibinfo{person}{Thomas Reitmaier}, \bibinfo{person}{Anja Venter}, {and} \bibinfo{person}{Edwin Blake}.} \bibinfo{year}{2016}\natexlab{}.
\newblock \showarticletitle{Enough with `In-The-Wild'}. In \bibinfo{booktitle}{\emph{Proceedings of the First African Conference on Human Computer Interaction}}. \bibinfo{pages}{182--186}.
\newblock


\bibitem[Stern(2020)]%
        {stern2020large}
\bibfield{author}{\bibinfo{person}{David~I Stern}.} \bibinfo{year}{2020}\natexlab{}.
\newblock \showarticletitle{How large is the economy-wide rebound effect?}
\newblock \bibinfo{journal}{\emph{Energy Policy}}  \bibinfo{volume}{147} (\bibinfo{year}{2020}), \bibinfo{pages}{111870}.
\newblock


\bibitem[Stevens et~al\mbox{.}(2019)]%
        {stevens2019using}
\bibfield{author}{\bibinfo{person}{Gunnar Stevens}, \bibinfo{person}{Paul Bossauer}, \bibinfo{person}{Stephanie Vonholdt}, {and} \bibinfo{person}{Christina Pakusch}.} \bibinfo{year}{2019}\natexlab{}.
\newblock \showarticletitle{Using time and space efficiently in driverless cars: Findings of a co-design study}. In \bibinfo{booktitle}{\emph{Proceedings of the 2019 CHI Conference on Human Factors in Computing Systems}}. \bibinfo{pages}{1--14}.
\newblock


\bibitem[Strengers(2014)]%
        {strengers2014smart}
\bibfield{author}{\bibinfo{person}{Yolande Strengers}.} \bibinfo{year}{2014}\natexlab{}.
\newblock \showarticletitle{Smart energy in everyday life: are you designing for resource man?}
\newblock \bibinfo{journal}{\emph{interactions}} \bibinfo{volume}{21}, \bibinfo{number}{4} (\bibinfo{year}{2014}), \bibinfo{pages}{24--31}.
\newblock


\bibitem[Strengers et~al\mbox{.}(2020)]%
        {strengers2020pursuing}
\bibfield{author}{\bibinfo{person}{Yolande Strengers}, \bibinfo{person}{Mike Hazas}, \bibinfo{person}{Larissa Nicholls}, \bibinfo{person}{Jesper Kjeldskov}, {and} \bibinfo{person}{Mikael~B Skov}.} \bibinfo{year}{2020}\natexlab{}.
\newblock \showarticletitle{Pursuing pleasance: Interrogating energy-intensive visions for the smart home}.
\newblock \bibinfo{journal}{\emph{International Journal of Human-Computer Studies}}  \bibinfo{volume}{136} (\bibinfo{year}{2020}), \bibinfo{pages}{102379}.
\newblock


\bibitem[Strengers et~al\mbox{.}(2019)]%
        {strengers2019protection}
\bibfield{author}{\bibinfo{person}{Yolande Strengers}, \bibinfo{person}{Jenny Kennedy}, \bibinfo{person}{Paula Arcari}, \bibinfo{person}{Larissa Nicholls}, {and} \bibinfo{person}{Melissa Gregg}.} \bibinfo{year}{2019}\natexlab{}.
\newblock \showarticletitle{Protection, productivity and pleasure in the smart home: emerging expectations and gendered insights from Australian early adopters}. In \bibinfo{booktitle}{\emph{Proceedings of the 2019 CHI conference on human factors in computing systems}}. \bibinfo{pages}{1--13}.
\newblock


\bibitem[Strengers et~al\mbox{.}(2016)]%
        {strengers2016hidden}
\bibfield{author}{\bibinfo{person}{Yolande Strengers}, \bibinfo{person}{Janine Morley}, \bibinfo{person}{Larissa Nicholls}, {and} \bibinfo{person}{Mike Hazas}.} \bibinfo{year}{2016}\natexlab{}.
\newblock \showarticletitle{The hidden energy costs of smart homes}.
\newblock \bibinfo{journal}{\emph{The conversation}}  \bibinfo{volume}{12} (\bibinfo{year}{2016}).
\newblock


\bibitem[Strengers(2011)]%
        {Strengers2011}
\bibfield{author}{\bibinfo{person}{Yolande~A.A. Strengers}.} \bibinfo{year}{2011}\natexlab{}.
\newblock \showarticletitle{Designing Eco-Feedback Systems for Everyday Life}. In \bibinfo{booktitle}{\emph{Proceedings of the SIGCHI Conference on Human Factors in Computing Systems}} (Vancouver, BC, Canada) \emph{(\bibinfo{series}{CHI '11})}. \bibinfo{publisher}{Association for Computing Machinery}, \bibinfo{address}{New York, NY, USA}, \bibinfo{pages}{2135–2144}.
\newblock
\showISBNx{9781450302289}
\urldef\tempurl%
\url{https://doi.org/10.1145/1978942.1979252}
\showDOI{\tempurl}


\bibitem[Tirado~Herrero et~al\mbox{.}(2018)]%
        {tirado_herrero_smart_2018}
\bibfield{author}{\bibinfo{person}{Sergio Tirado~Herrero}, \bibinfo{person}{Larissa Nicholls}, {and} \bibinfo{person}{Yolande Strengers}.} \bibinfo{year}{2018}\natexlab{}.
\newblock \showarticletitle{Smart home technologies in everyday life: do they address key energy challenges in households?}
\newblock \bibinfo{journal}{\emph{Current Opinion in Environmental Sustainability}}  \bibinfo{volume}{31} (\bibinfo{date}{April} \bibinfo{year}{2018}), \bibinfo{pages}{65--70}.
\newblock
\showISSN{18773435}
\urldef\tempurl%
\url{https://doi.org/10.1016/j.cosust.2017.12.001}
\showDOI{\tempurl}


\bibitem[Tsao et~al\mbox{.}(2010)]%
        {tsao_solid-state_2010}
\bibfield{author}{\bibinfo{person}{J.~Y. Tsao}, \bibinfo{person}{H.~D. Saunders}, \bibinfo{person}{J.~R. Creighton}, \bibinfo{person}{M.~E. Coltrin}, {and} \bibinfo{person}{J.~A. Simmons}.} \bibinfo{year}{2010}\natexlab{}.
\newblock \showarticletitle{Solid-state lighting: an energy-economics perspective}.
\newblock \bibinfo{journal}{\emph{Journal of Physics D: Applied Physics}} \bibinfo{volume}{43}, \bibinfo{number}{35} (\bibinfo{date}{Aug.} \bibinfo{year}{2010}), \bibinfo{pages}{354001}.
\newblock
\showISSN{0022-3727}
\urldef\tempurl%
\url{https://doi.org/10.1088/0022-3727/43/35/354001}
\showDOI{\tempurl}
\newblock
\shownote{Publisher: IOP Publishing}.


\bibitem[Turner(2009)]%
        {turner2009negative}
\bibfield{author}{\bibinfo{person}{Karen Turner}.} \bibinfo{year}{2009}\natexlab{}.
\newblock \showarticletitle{Negative rebound and disinvestment effects in response to an improvement in energy efficiency in the UK economy}.
\newblock \bibinfo{journal}{\emph{Energy Economics}} \bibinfo{volume}{31}, \bibinfo{number}{5} (\bibinfo{year}{2009}), \bibinfo{pages}{648--666}.
\newblock


\bibitem[Turner(2013)]%
        {turner2013rebound}
\bibfield{author}{\bibinfo{person}{Karen Turner}.} \bibinfo{year}{2013}\natexlab{}.
\newblock \showarticletitle{``Rebound'' effects from increased energy efficiency: a time to pause and reflect}.
\newblock \bibinfo{journal}{\emph{The Energy Journal}} \bibinfo{volume}{34}, \bibinfo{number}{4} (\bibinfo{year}{2013}).
\newblock


\bibitem[{United Nations}(2022)]%
        {UNGoalsreport}
\bibfield{author}{\bibinfo{person}{{United Nations}}.} \bibinfo{year}{2022}\natexlab{}.
\newblock \bibinfo{title}{The Sustainable Development Goals Report 2022}.
\newblock
\newblock


\bibitem[{United Nations Environment Programme}(2021)]%
        {UNreport}
\bibfield{author}{\bibinfo{person}{{United Nations Environment Programme}}.} \bibinfo{year}{2021}\natexlab{}.
\newblock \bibinfo{title}{2021 Global Status Report for Buildings and Construction: Towards a Zero‑emission, Efficient and Resilient Buildings and Construction Sector}.
\newblock
\newblock


\bibitem[Vadén et~al\mbox{.}(2020)]%
        {vaden_decoupling_2020}
\bibfield{author}{\bibinfo{person}{T. Vadén}, \bibinfo{person}{V. Lähde}, \bibinfo{person}{A. Majava}, \bibinfo{person}{P. Järvensivu}, \bibinfo{person}{T. Toivanen}, \bibinfo{person}{E. Hakala}, {and} \bibinfo{person}{J.~T. Eronen}.} \bibinfo{year}{2020}\natexlab{}.
\newblock \showarticletitle{Decoupling for ecological sustainability: {A} categorisation and review of research literature}.
\newblock \bibinfo{journal}{\emph{Environmental Science \& Policy}}  \bibinfo{volume}{112} (\bibinfo{date}{Oct.} \bibinfo{year}{2020}), \bibinfo{pages}{236--244}.
\newblock
\showISSN{1462-9011}
\urldef\tempurl%
\url{https://doi.org/10.1016/j.envsci.2020.06.016}
\showDOI{\tempurl}


\bibitem[Van~Thillo et~al\mbox{.}(2022)]%
        {van_thillo_potential_2022}
\bibfield{author}{\bibinfo{person}{L. Van~Thillo}, \bibinfo{person}{S. Verbeke}, {and} \bibinfo{person}{A. Audenaert}.} \bibinfo{year}{2022}\natexlab{}.
\newblock \showarticletitle{The potential of building automation and control systems to lower the energy demand in residential buildings: {A} review of their performance and influencing parameters}.
\newblock \bibinfo{journal}{\emph{Renewable and Sustainable Energy Reviews}}  \bibinfo{volume}{158} (\bibinfo{year}{2022}), \bibinfo{pages}{112099}.
\newblock
\showISSN{13640321}
\urldef\tempurl%
\url{https://doi.org/10.1016/j.rser.2022.112099}
\showDOI{\tempurl}


\bibitem[Walzberg et~al\mbox{.}(2020)]%
        {walzberg_should_2020}
\bibfield{author}{\bibinfo{person}{Julien Walzberg}, \bibinfo{person}{Thomas Dandres}, \bibinfo{person}{Nicolas Merveille}, \bibinfo{person}{Mohamed Cheriet}, {and} \bibinfo{person}{Réjean Samson}.} \bibinfo{year}{2020}\natexlab{}.
\newblock \showarticletitle{Should we fear the rebound effect in smart homes?}
\newblock \bibinfo{journal}{\emph{Renewable and Sustainable Energy Reviews}}  \bibinfo{volume}{125} (\bibinfo{date}{June} \bibinfo{year}{2020}), \bibinfo{pages}{109798}.
\newblock
\showISSN{13640321}
\urldef\tempurl%
\url{https://doi.org/10.1016/j.rser.2020.109798}
\showDOI{\tempurl}


\bibitem[Wang et~al\mbox{.}(2018)]%
        {wang2018evaluation}
\bibfield{author}{\bibinfo{person}{Xiaolei Wang}, \bibinfo{person}{Xiaohui Wen}, {and} \bibinfo{person}{Chunping Xie}.} \bibinfo{year}{2018}\natexlab{}.
\newblock \showarticletitle{An evaluation of technical progress and energy rebound effects in China's iron \& steel industry}.
\newblock \bibinfo{journal}{\emph{Energy Policy}}  \bibinfo{volume}{123} (\bibinfo{year}{2018}), \bibinfo{pages}{259--265}.
\newblock


\bibitem[Wang et~al\mbox{.}(2016)]%
        {wang_measurement_2016}
\bibfield{author}{\bibinfo{person}{Zhaohua Wang}, \bibinfo{person}{Bai Han}, {and} \bibinfo{person}{Milin Lu}.} \bibinfo{year}{2016}\natexlab{}.
\newblock \showarticletitle{Measurement of energy rebound effect in households: {Evidence} from residential electricity consumption in {Beijing}, {China}}.
\newblock \bibinfo{journal}{\emph{Renewable and Sustainable Energy Reviews}}  \bibinfo{volume}{58} (\bibinfo{date}{May} \bibinfo{year}{2016}), \bibinfo{pages}{852--861}.
\newblock
\showISSN{13640321}
\urldef\tempurl%
\url{https://doi.org/10.1016/j.rser.2015.12.179}
\showDOI{\tempurl}


\bibitem[Watson and Kharrufa(2021)]%
        {watson2021hci}
\bibfield{author}{\bibinfo{person}{Colin Watson} {and} \bibinfo{person}{Ahmed Kharrufa}.} \bibinfo{year}{2021}\natexlab{}.
\newblock \showarticletitle{HCI-H is also for Hazard: Using HAZOP to Identify Undesirable Consequences in Socio-Technical Systems}. In \bibinfo{booktitle}{\emph{ACM SIGCAS Conference on Computing and Sustainable Societies}}. \bibinfo{pages}{230--242}.
\newblock


\bibitem[Widdicks et~al\mbox{.}(2019)]%
        {widdicks2019streaming}
\bibfield{author}{\bibinfo{person}{Kelly Widdicks}, \bibinfo{person}{Mike Hazas}, \bibinfo{person}{Oliver Bates}, {and} \bibinfo{person}{Adrian Friday}.} \bibinfo{year}{2019}\natexlab{}.
\newblock \showarticletitle{Streaming, multi-screens and YouTube: The new (unsustainable) ways of watching in the home}. In \bibinfo{booktitle}{\emph{Proceedings of the 2019 CHI Conference on Human Factors in Computing Systems}}. \bibinfo{pages}{1--13}.
\newblock


\bibitem[Widdicks et~al\mbox{.}(2023)]%
        {widdicks2023:rebound-systems}
\bibfield{author}{\bibinfo{person}{Kelly Widdicks}, \bibinfo{person}{Federica Lucivero}, \bibinfo{person}{Gabrielle Samuel}, \bibinfo{person}{Lucas~Somavilla Croxatto}, \bibinfo{person}{Marcia~Tavares Smith}, \bibinfo{person}{Carolyn~Ten Holter}, \bibinfo{person}{Mike Berners-Lee}, \bibinfo{person}{Gordon~S. Blair}, \bibinfo{person}{Marina Jirotka}, \bibinfo{person}{Bran Knowles}, \bibinfo{person}{Steven Sorrell}, \bibinfo{person}{Miriam~Börjesson Rivera}, \bibinfo{person}{Caroline Cook}, \bibinfo{person}{Vlad~C. Coroamă}, \bibinfo{person}{Timothy~J. Foxon}, \bibinfo{person}{Jeffrey Hardy}, \bibinfo{person}{Lorenz~M. Hilty}, \bibinfo{person}{Simon Hinterholzer}, {and} \bibinfo{person}{Birgit Penzenstadler}.} \bibinfo{year}{2023}\natexlab{}.
\newblock \showarticletitle{Systems thinking and efficiency under emissions constraints: Addressing rebound effects in digital innovation and policy}.
\newblock \bibinfo{journal}{\emph{Patterns}} \bibinfo{volume}{4}, \bibinfo{number}{2} (\bibinfo{date}{Feb.} \bibinfo{year}{2023}).
\newblock
\showISSN{2666-3899}
\urldef\tempurl%
\url{https://doi.org/10.1016/j.patter.2023.100679}
\showDOI{\tempurl}


\bibitem[Wiedmann et~al\mbox{.}(2015)]%
        {wiedmann_material_2015}
\bibfield{author}{\bibinfo{person}{Thomas~O. Wiedmann}, \bibinfo{person}{Heinz Schandl}, \bibinfo{person}{Manfred Lenzen}, \bibinfo{person}{Daniel Moran}, \bibinfo{person}{Sangwon Suh}, \bibinfo{person}{James West}, {and} \bibinfo{person}{Keiichiro Kanemoto}.} \bibinfo{year}{2015}\natexlab{}.
\newblock \showarticletitle{The material footprint of nations}.
\newblock \bibinfo{journal}{\emph{Proceedings of the National Academy of Sciences}} \bibinfo{volume}{112}, \bibinfo{number}{20} (\bibinfo{date}{May} \bibinfo{year}{2015}), \bibinfo{pages}{6271--6276}.
\newblock
\urldef\tempurl%
\url{https://doi.org/10.1073/pnas.1220362110}
\showDOI{\tempurl}
\newblock
\shownote{Publisher: Proceedings of the National Academy of Sciences}.


\bibitem[Williams(2011)]%
        {williams2011:indirect-effects}
\bibfield{author}{\bibinfo{person}{Eric Williams}.} \bibinfo{year}{2011}\natexlab{}.
\newblock \showarticletitle{Environmental effects of information and communications technologies}.
\newblock \bibinfo{journal}{\emph{Nature}} \bibinfo{volume}{479}, \bibinfo{number}{7373} (\bibinfo{date}{Nov.} \bibinfo{year}{2011}), \bibinfo{pages}{354--358}.
\newblock
\showISSN{0028-0836, 1476-4687}
\urldef\tempurl%
\url{https://doi.org/10.1038/nature10682}
\showDOI{\tempurl}


\bibitem[Williams and Matthews(2007)]%
        {williams_scoping_2007}
\bibfield{author}{\bibinfo{person}{Eric~D Williams} {and} \bibinfo{person}{H~Scott Matthews}.} \bibinfo{year}{2007}\natexlab{}.
\newblock \showarticletitle{Scoping the potential of monitoring and control technologies to reduce energy use in homes}. In \bibinfo{booktitle}{\emph{Proceedings of the 2007 IEEE International Symposium on Electronics and the Environment}}. IEEE, \bibinfo{pages}{239--244}.
\newblock


\end{thebibliography}

\appendix

\end{document}